\documentclass[twocolumn]{aastex631}
\usepackage[utf8]{inputenc}
\usepackage{hyperref}
\usepackage{bm}
\hypersetup{
    colorlinks=true,
    filecolor=magenta,      
    urlcolor=blue,
}
\usepackage{float}

\newcommand{\kms}{km s$^{-1}$}
\newcommand{\Mjsr}{MJy sr$^{-1}$}

\newcommand{\jybe}{Jy beam$^{-1}$}
\newcommand{\hii}{H~{\scriptsize II}}

\newcommand{\cm}{cm$^{-3}$}

\newcommand{\msuny}{M$_{\odot}$ yr$^{-1}$}

\newcommand{\til}{$\sim$}
\newcommand{\ring}{M0.8--0.2}
\newcommand{\ie}{i.e.}
\newcommand{\eg}{e.g.}

\newcommand{\degree}{$^{\circ}$}

\newcommand{\h}{$^{\mathrm{h}}$}
\newcommand{\m}{$^{\mathrm{m}}$}
\newcommand{\s}{$^{\mathrm{s}}$}

\begin{document}

\title{\uppercase{SOFIA/HAWC+ Far-Infrared Polarimetric Large Area CMZ Exploration (FIREPLACE) Survey I: General Results from the Pilot Program } }

\author[0000-0002-4013-6469]{Natalie O. Butterfield}
\affiliation{National Radio Astronomy Observatory, 520 Edgemont Road, Charlottesville, VA 22903, USA}
\email{nbutterf@nrao.edu}
\affiliation{Department of Physics, Villanova University, 800 E. Lancaster Ave., Villanova, PA 19085, USA}
\author[0000-0003-0016-0533]{David T. Chuss}
\affil{Department of Physics, Villanova University, 800 E. Lancaster Ave., Villanova, PA 19085, USA}
\author[0000-0001-8819-9648]{Jordan A. Guerra}
\affil{Department of Physics, Villanova University, 800 E. Lancaster Ave., Villanova, PA 19085, USA}
\author[0000-0002-6753-2066]{Mark R. Morris}
\affil{Department of Physics and Astronomy, University of California, Los Angeles, Box 951547, Los Angeles, CA 90095-1547 USA}
\author[0000-0002-5811-0136]{Dylan Par\'e}
\affiliation{Department of Physics, Villanova University, 800 E. Lancaster Ave., Villanova, PA 19085, USA}
\author[0000-0002-7567-4451]{Edward J. Wollack}
\affil{NASA Goddard Space Flight Center, Greenbelt, MD 20771, USA}
\author{C. Darren Dowell}
\affil{NASA Jet Propulsion Laboratory, California Institute of Technology, 4800 Oak Grove Drive, Pasadena, CA 91109, USA}
\author[0000-0001-9315-8437]{Matthew J. Hankins}
\affiliation{Arkansas Tech University, 215 West O Street, Russellville, AR 72801, USA}
\author[0009-0006-4830-163X]{Kaitlyn Karpovich}
\affiliation{Department of Physics, Villanova University, 800 E. Lancaster Ave., Villanova, PA 19085, USA}
\author[0000-0001-5389-5635]{Javad Siah}
\affiliation{Department of Physics, Villanova University, 800 E. Lancaster Ave., Villanova, PA 19085, USA}
\author[0000-0002-8437-0433]{Johannes Staguhn}
\affil{NASA Goddard Space Flight Center, Greenbelt, MD 20771, USA}
\affiliation{Department of Physics and Astronomy, Johns Hopkins University, 3400 North Charles Street, Baltimore, MD, 21218, USA}
\author[0000-0003-4821-713X]{Ellen Zweibel}
\affiliation{Department of Astronomy, U. Wisconsin-Madison, 475 N Charter Street, Madison, WI 53706, USA}

\begin{abstract}

We present the first data release (DR1) of the Far-Infrared Polarimetric Large Area CMZ Exploration (FIREPLACE) survey. The survey was taken using the 214-\micron\ band of the HAWC+ instrument with the SOFIA telescope (19\farcs6 resolution; 0.7 pc).  In this first data release we present dust polarization observations covering a \til0.5\degree\ region of the Galactic Center's Central Molecular Zone (CMZ), approximately centered on the Sgr B2 complex. We detect \til25,000 Nyquist-sampled polarization pseudovectors, after applying {the standard SOFIA} cuts for minimum signal-to-noise in fractional polarization and total intensity of 3 and 200, respectively.  Analysis of the magnetic field orientation suggests a bimodal distribution in the field direction. This bimodal distribution shows enhancements in the distribution of field directions for orientations parallel and perpendicular to the Galactic plane, which is suggestive of a CMZ magnetic field configuration with polodial and torodial components. Furthermore, a detailed analysis of individual clouds included in our survey (i.e., Sgr B2, Sgr B2-NW, Sgr B2-Halo, Sgr B1, and Clouds-E/F) shows these clouds have fractional polarization values of 1--10\% at 214-\micron, with most of the emission having values $<$5\%. A few of these clouds (i.e., Sgr B2, Clouds-E/F) show relatively low fractional polarization values toward the cores of the cloud, with higher fractional polarization values toward the less dense periphery. We also observe higher fractional polarization towards compact \hii\ regions which could indicate an enhancement in the grain alignment in the dust surrounding these sources.

\end{abstract}


\section{Introduction}
\label{intro}

\begin{figure*}
\centering
\includegraphics[width=1.0\textwidth]{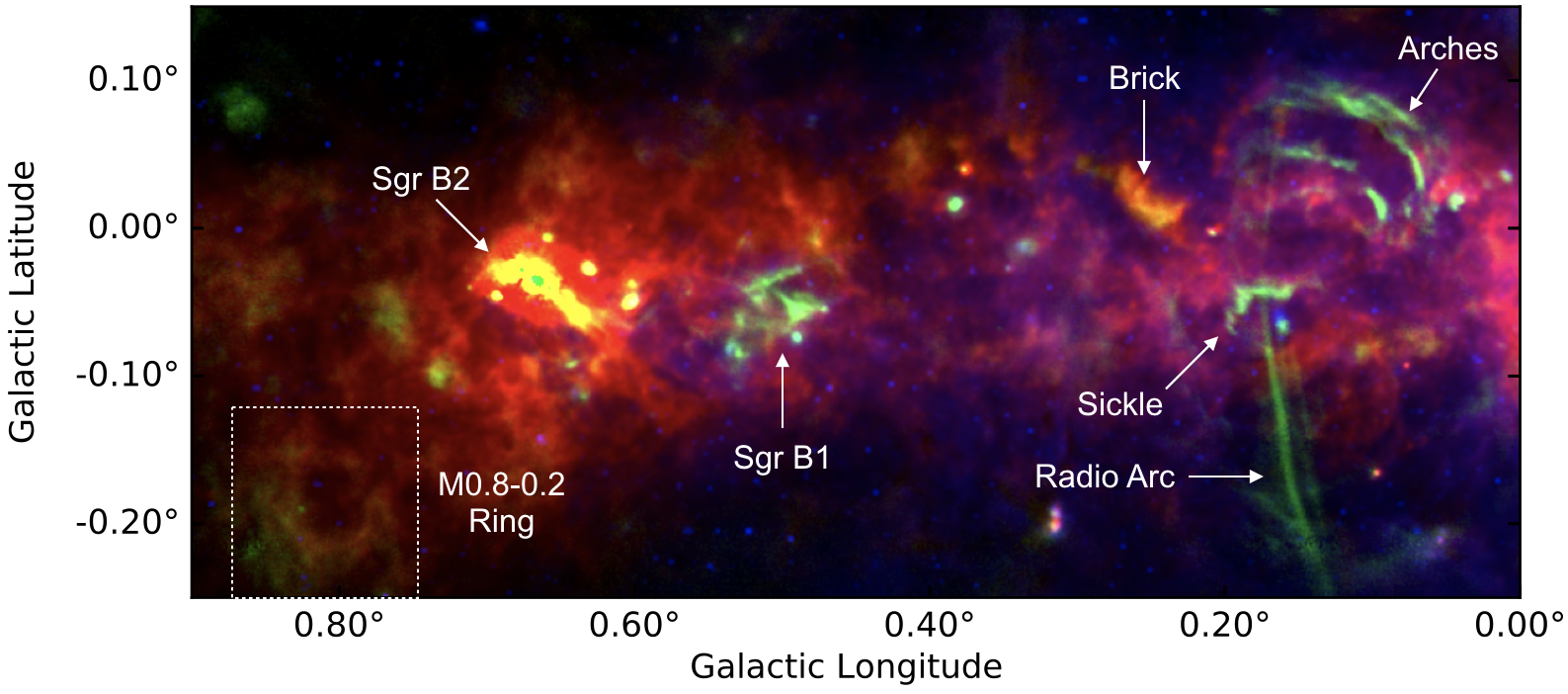}
\caption{Three-color image of the Eastern 125 parsecs of the Galactic Center. This image shows the far-infrared, 160 \micron\ emission from Herschel in red \citep[Hi-GAL survey, PACS instrument, 12\arcsec\ angular resolution;][]{molinari16}, the GBT+MUSTANG 3-mm (90 GHz) microwave emission in green \citep[Mustang Galactic Plane Survey (MGPS), 9\arcsec\ angular resolution;][]{ginsburg20}, and the 8 \micron\ emission from Spitzer in blue \citep[GLIMPSE II survey, IRAC4 band, 1.2\arcsec\ angular resolution;][]{Churchwell09}. Labeled are several prominent features in the CMZ. The white dashed box at the far left shows the location of the \ring\ cloud (FIREPLACE II, Butterfield et al. 2023, submitted). 
}
\label{3color}
\end{figure*}

The Central Molecular Zone (CMZ) of the Milky Way's Galactic Center (GC), at a distance of 8.2 kpc from Earth \citep{gravity}, is an extreme environment compared with the local interstellar medium (ISM). The molecular gas in this inner region of the Galaxy is denser \citep[10$^3$--10$^6$ \cm; \eg,][]{mills18}, hotter \citep[50$-$300 K; \eg,][]{krieger17} and more turbulent \citep[10 \kms; \eg,][]{kauffmann17a} than gas in the disk of the Milky Way Galaxy. The magnetic fields in this region of the Galaxy are also stronger than fields in the disk of the Galaxy, reaching strengths in the $\sim$mG range \citep[\eg,][]{zadeh87,chuss03, pillai15, Mangilli19}. {These magnetic field values are heavily debated, with some radio observations implying mean values of 100--400 $\mu$G \citep{zadeh22}, or as low as a few $\mu$G \citep{LaRosa05}. However, these measurements apply to scales of hundreds of parsecs.  Such estimates assume equipartition and therefore could represent a lower limit \citep[\eg,][see their Section 3]{morris06}. }

Clouds of dense gas and dust in the CMZ are organized into a large \til200 pc `twisted ring' structure \citep[\eg,][]{molinari11}. The dense gas and dust are concentrated largely into orbiting streams that might have a ring morphology, a spiral arm morphology, or an open stream morphology \citep[\eg,][]{sofue95,molinari11,Kru15, Henshaw16}, in which individual clouds all tend to follow a similar orbital path. Figure \ref{3color} shows a 3 color image of the Eastern 125 pc of the CMZ. In this figure the dust tracing the dense molecular clouds, is shown in red, the thermal Bremsstrahlung (free-free; e.g., Sgr B2, Sickle), bright non-thermal Synchrotron emission (e.g., Radio Arc), and colder dust is shown in green, and the PAH emission tracing high-mass stars is in blue. The Eastern region of the CMZ contains numerous dense molecular clouds, the well-known massive star-forming complex, Sgr B2, and several magnetic field structures that are bright at radio wavelengths (e.g., Radio Arc), also known as non-thermal filaments (NTFs).

NTFs are perhaps the most striking manifestations of magnetic fields in the central 100 pc of the Milky Way and some of the first sources utilized to study magnetic fields in the CMZ.  NTFs are long, straight structures along which relativistic electrons propagate and trace the magnetic field via Synchrotron emission. The Radio Arc (annotated in Figure \ref{3color}) was the first of these structures discovered \citep{zadeh1984}; however, since then many such structures {have been observed}, mostly oriented perpendicular to the Galactic plane \citep[\eg,][]{LaRosa04,heywood22, zadeh22}. Since the first detection of the Radio Arc in 1984, magnetic fields in the CMZ {have been studied} numerous times using a variety of methods \citep[\eg,][]{zadeh87, novak00, nishiyama10, guan21}.

Magnetic fields in the CMZ {have also been explored} at radio wavelengths using rotation measure (RM) synthesis techniques at radio wavelengths \citep[\til1--10 GHz; \eg,][]{zadeh87, lang99, Law+11, zadeh22, Pare21}.  The magnetic field {geometry} in the CMZ {has also been characterized and interpreted in the context of} magnetically aligned dust grains in the near-infrared via extinction polarimetry \citep[\textit{J} (1.25~\micron), \textit{H} (1.63~\micron), and \textit{K$_s$} (2.14~\micron) bands; e.g.,][]{nishiyama09,nishiyama10} and in the far-infrared \citep[{50 \micron\ -- 450 \micron; \eg,}][]{Hildebrand+93,novak00,Novak2003,chuss03,Mangilli19,Guerra23} and microwave \citep[{1.3 mm -- 3.3 mm, 90--220 GHz;}][]{guan21} via dust emission polarimetry. Recent far-infrared (PILOT, 2\farcm2) and microwave (ACTpol, 1--2\arcmin) surveys of large-scale gas structures (2--3 pc), using observations of polarized dust emission, show an organized {magnetic} field with a 22\degree\ tilt in the magnetic field relative to the Galactic Plane \citep[][respectively]{Mangilli19, guan21}. {The magnetic field directions inferred from ACTPOL and PILOT polarimetric observations are oriented \til20\degree\ North of East \citep[see Figure 1, top, in][for a visual of this orientation in the PILOT dataset]{Mangilli19}.}
This magnetic field orientation is consistent with near-IR observations \citep[e.g.,][]{nishiyama09,nishiyama10}. A comparison of the polarization between stars on the near and far sides of the CMZ suggests that this \til20\degree\ polarization orientation is associated with material in the inner 1$-$2 kpc of the Milky Way galaxy \citep[][]{nishiyama09}. This tension between the general field direction observed in ionized versus molecular material in the Galactic center {presents an apparent contradiction, the resolution of which may provide insight into the three-dimensional field structure of the region}. \citet{Novak2003} suggest a paradigm in which an initially-poloidal magnetic field is sheared into a toroidal configuration in the denser molecular regions of the GC. 

Dust polarization observations on the size scales of individual clouds in the CMZ \citep[\eg,][]{Dotson00, novak00, Novak2003, chuss03, Dotson10, pillai15} show that the fields tracing the clouds are more consistent with the local cloud morphology than they are with the large-scale trend in the field direction. For example, \cite{chuss03} and \citet{novak00} used the CSO telescope to observe dust polarization at 350 \micron\ (20\arcsec\ angular resolution) in seven clouds in the central 50 pc of the CMZ. They observed that the field direction generally followed the cloud morphology on these size scales \citep[][see their Figure 3]{chuss03}.

\begin{figure*}
    \centering 
    \includegraphics[width=1.0\textwidth]{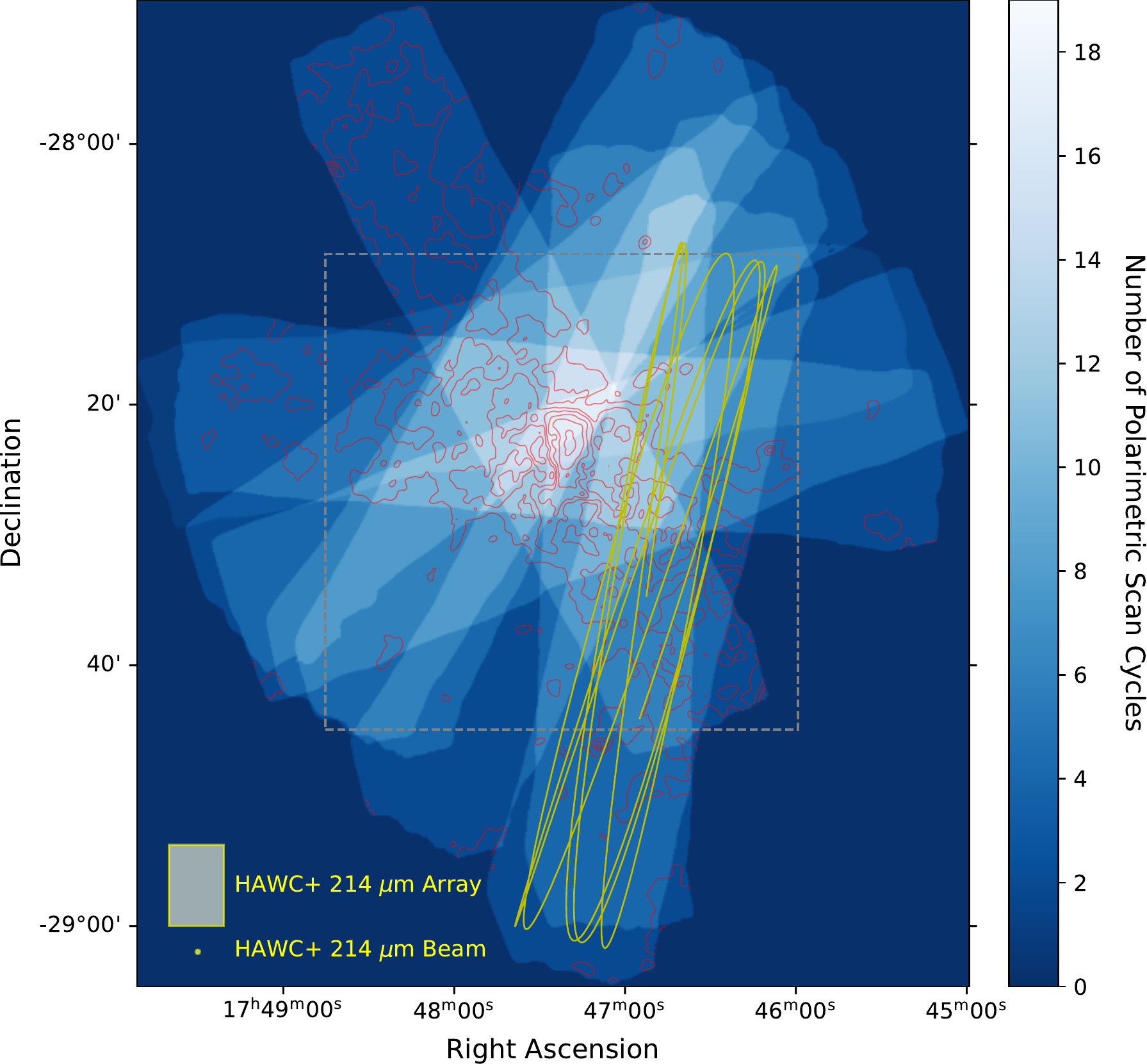} 
    \caption{Coverage of the FIREPLACE pilot study. The elongated rectangular patches show the extent of each scan where the brightness of each pixel in the map indicates the total number of polarimetric scans during which that pixel was covered. The dashed gray box shows the field of view for Figures \ref{85rounds-crop}, \ref{fig:LICs}, \ref{hist-fig-map}, \ref{pol-fig}, and \ref{85rounds-lic}, which contain the majority of the CMZ emission in this region and includes the highest cross-referenced area. Contours of the HAWC+ Stokes $I$ data are shown in red.  A representative Lissajous path for the center of the array is shown for one of these scan patterns (yellow path).  At the bottom of the image, an array footprint and the beam size are shown. 
    }
    \label{hitmap}
\end{figure*}

In this paper we present the first data release of a recent SOFIA/HAWC+ Legacy Survey of dust polarization in the CMZ: the SOFIA/HAWC+ Far-Infrared Polarimetric Large Area CMZ Exploration (FIREPLACE) Survey. The FIREPLACE pilot survey covered \til0.5\degree\ of the CMZ at 19\farcs6 angular resolution (0.7 pc at the Galactic Center, comparable to the size scales of small clouds in the CMZ). We present an overview of the current observations of the SOFIA legacy survey in Section \ref{survey} and investigate the contrast between the large and small-scale dust polarization studies in Section \ref{Magnetosphere}. We investigate the magnetic field orientation across the DR1 sample in Section \ref{sec:histogram} and then focus in section \ref{sec:sources} on a few notable clouds included in our survey (e.g., Sgr B2, etc). Lastly, we summarize our conclusions for the first data release of the FIREPLACE survey in Section \ref{conclusion}. Additional data reduction methodology is also presented in the Appendix.

\begin{figure*}
\centering
\includegraphics[width=0.95\textwidth]{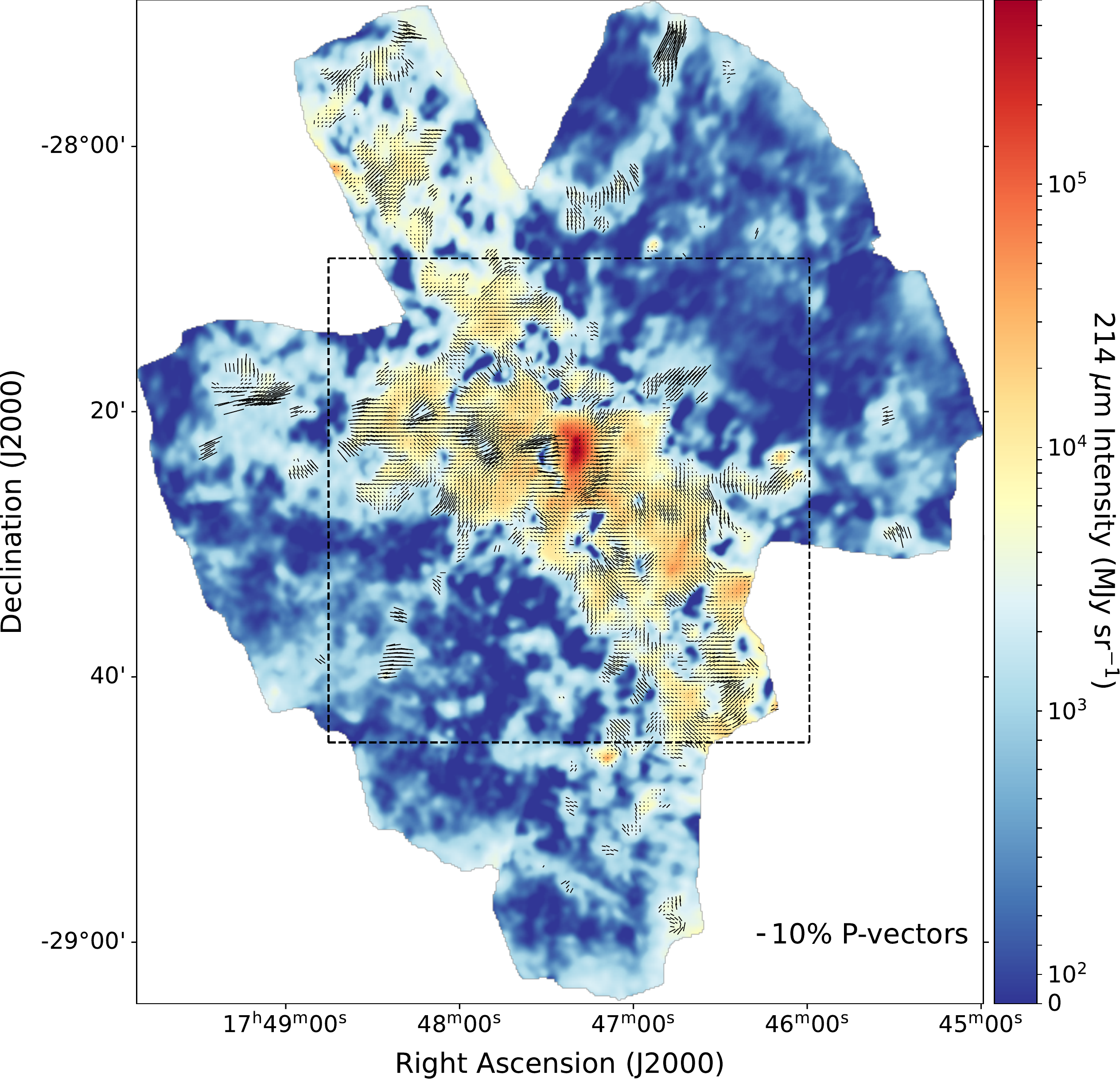}
\caption{Full map of the first data release for the 214 \micron\ FIREPLACE Legacy Survey, showing the {derived electric field (polarization direction) pseudovectors, using equations \ref{eq:phi} and \ref{eq:p},} for the entirety of the mapped region. The data reduction steps used to produce this image are outlined in Section~\ref{obs}. The polarization cuts for the magnetic field pseudovectors shown are the HAWC+ defaults: $p$/$\sigma_p$$>$3.0, $p$$<$50\%, $I_{peak}$$>$0.0, and $I$/$\sigma_I$$>$200. The black dashed box shows the region with the best cross-linking between scans, from Figure \ref{hitmap} (gray box), and which corresponds to the field-of-view in Figures \ref{85rounds-crop}, \ref{fig:LICs}, \ref{hist-fig-map}, \ref{pol-fig}, and \ref{85rounds-lic}. {A 10\% fractional polarization pseudovector is shown in the bottom right corner, for reference.}
}
\label{85rounds}
\end{figure*} 

\section{FIREPLACE Survey Pilot Program}
\label{survey}

\subsection{Overview of the Legacy Survey}
\label{survey2}

The goal of the FIREPLACE Legacy Program is to measure the polarization throughout the CMZ at 214 \micron\ to characterize the magnetic field in the cool dust component that has been observed in total intensity at 250 \micron\ by the Herschel Observatory \citep{molinari10}. This paper reports on the initial pilot program for this Legacy Survey.

The High Angular Resolution Wideband Camera+ \citep[HAWC+;][]{Harper2018} on the Stratospheric Observatory for Infrared Astronomy (SOFIA) provides polarimetric imaging at four bands in the far-infrared. FIREPLACE utilizes the 214 \micron\ band, which has an effective angular resolution of 19\farcs6, a bandwidth of 44 \micron\ and an instantaneous field-of-view for polarimetry of 4\farcm2$\times$6\farcm2.

\subsection{Mapping Strategy}
\label{mapping}

In addition to the original chop-nod-match polarimetry mode, SOFIA has implemented \emph{on-the-fly mapping} (OTFMAP) polarimetry. The key advantage of OTFMAP polarimetry over the chop-nod-match mode is a significant increase in observatory efficiency. OTFMAP polarimetry entails rapid scanning of the array over the source. This scanning is repeated for each of four half-wave plate positions to fully and symmetrically sample the Stokes $Q$ and $U$ parameters. The HAWC+/SOFIA scan-mode polarimetry pipeline is based on the \texttt{CRUSH} \citep{Kovacs2008a} data processing software, which iteratively fits for correlated ``noise'' across the array as well as noise-based weights for the data and then produces a map in which the correlated noise signals are removed. In designing scan strategies around \texttt{CRUSH}, Lissajous patterns are chosen, because they enable a high degree of cross-linking in the process of sampling the scan, which mitigates systematic errors in image reconstruction \citep{Kovacs2008b}.

\begin{figure*}
\centering
\includegraphics[width=1.0\textwidth]{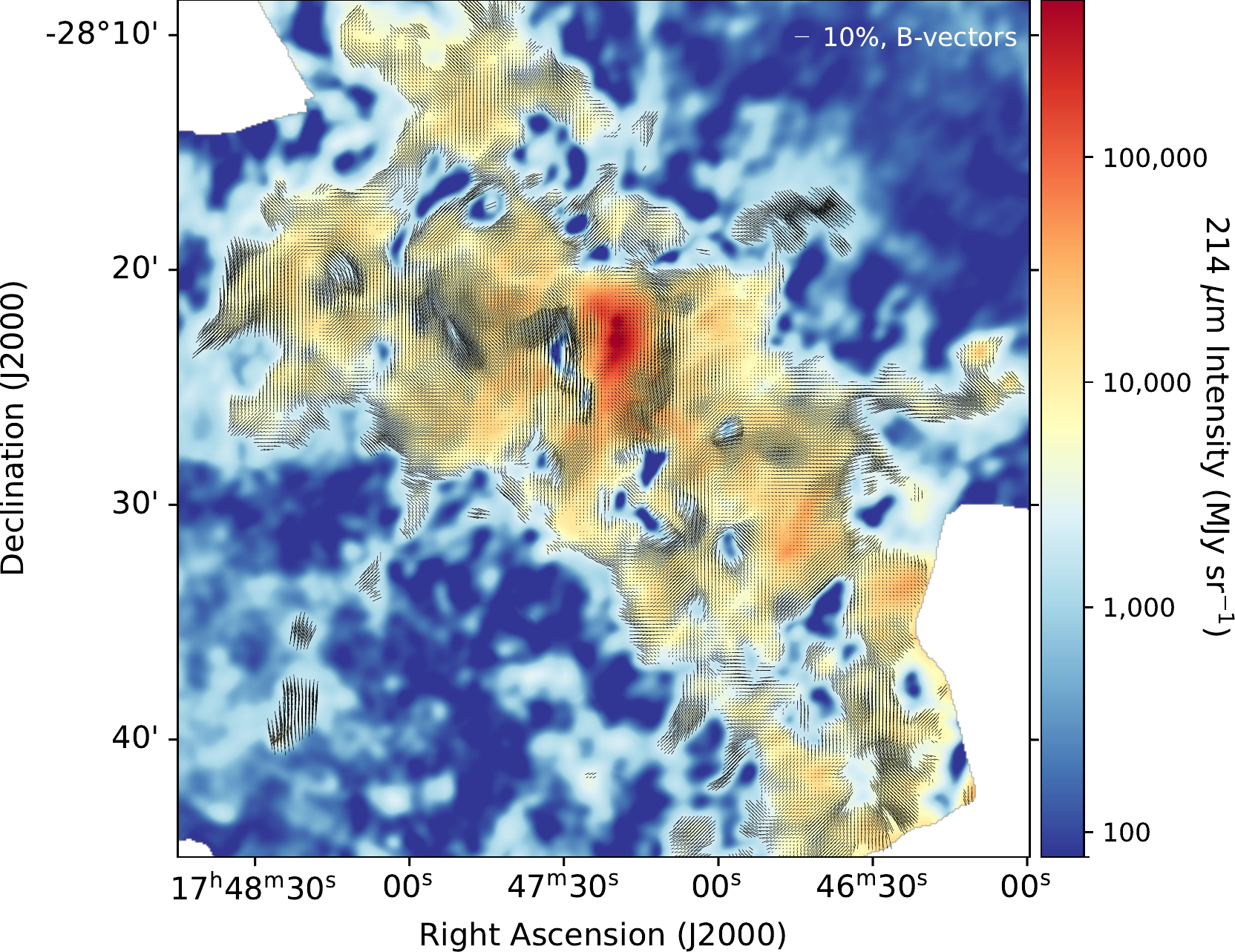}
\caption{
{Inferred magnetic field direction using the polarization pseudovectors from Figure \ref{85rounds} for the highly cross-linked region (dashed boxes in Figures \ref{hitmap} and \ref{85rounds}). For the inferred magnetic field directions we are assuming B-RAT alignment, as discussed in Section \ref{obs}, and have therefore rotated the polarization pseudovectors shown in Figure \ref{85rounds} by 90\degree. The length of the pseudovectors shown here is proportional to the fractional polarization. A 10\% fractional polarization pseudovector is shown in the top right corner, for reference.}
An LIC version of this figure is shown in Figure \ref{85rounds-lic}. 
}
\label{85rounds-crop}
\end{figure*}

The OTFMAP polarimetry capability for SOFIA/HAWC+ has been demonstrated for targets having spatial extents smaller than the instantaneous field-of-view of the detector \citep{Lopez-Rodriguez2022}. The CMZ presents a particular challenge, given its large size and substantial extended flux.  The key challenge here is that, for emission structures that span a larger angular size than the detector array, it is difficult for \texttt{CRUSH} to distinguish between these structures and correlated noise. However, repeated scans, particularly scans in different directions, can help differentiate between correlated noise and extended emission. Our strategy for FIREPLACE is to construct the large map from a series of OTFMAP polarimetry scans. We do so with elongated scans that, for the most part, cross the Galactic plane. This is done so that we can quickly scan from low intensity to high intensity regions to ensure that our defined baseline corresponds to the lower intensity regions off of the plane. The elongated scans allow us to revisit pixels faster than a two-dimensional scan with equal extent in both dimensions. We also try to scan at a variety of angles to mitigate the effect of any possible systematics associated with a particular scan direction.

\subsection{Observations and Data Reduction}
\label{obs}


{We highlight the observation strategy here. For further details on the data validation, the interested reader is referred to the Appendix (Section \ref{app-data-reduction}).} 
The data presented in this first data release (DR1) were obtained on two flights, F775 and F777, on August 31, 2021 and September 2, 2021, respectively. Both flights were initiated from Palmdale, CA, because COVID limitations prevented a Southern Hemisphere deployment in summer 2021. As such, the zenith angle for all observations was around 60$^\circ$. 

{Our map of the Eastern third of the CMZ is co-added from a series of strips.} We obtained a total of 10 strips. Each strip is {formed from the area swept out by the detector array as the boresight follows a Lissajous pattern. The scan has amplitudes of 27\arcmin and 4\farcm5 in the two directions and the ratio of frequencies between the two parametric sinusoidal curves in the Lissajous pattern is $\sim\sqrt{2}$, with the higher frequency corresponding to the long axis.} Each scan duration is 120 seconds, and four such scans are done for each polarimetric scan set, each corresponding to one of four half-wave plate angles (0$^\circ$, 22.5$^\circ$, 45$^\circ$, 67.5$^\circ$). This enables complete and uniform sampling of the $Q$ and $U$ Stokes parameters.

Nine of the strips were observed with two cycles of polarimetric scans for all four half-wave plate positions (\emph{i.e.}, eight scans in total for each strip).  One field was only observed once due to time limitations on Flight 775. Because of the novel nature of this observing mode, some scans were observed at scan angles that did not match those planned; however, these were included in the reduction nonetheless. Figure \ref{hitmap} shows the footprint of each scan and the number of complete polarization scan sets (1 set = 4 scans -- one for each of the 4 half-wave plate settings) during which each map pixel was observed. A representative Lissajous path for the center of the detector array for a single strip, the array footprint, and the beam size are also shown in Figure \ref{hitmap}. These observations cover roughly 0.5\degree\ of the CMZ, with a total on-sky integration time of 9,120 seconds (152 minutes) for the DR1 (pilot) dataset.

The observations were reduced with the HAWC+ data reduction pipeline (DRP 2.7.0) 
in \textit{scanpol} mode for each strip. We utilized the ``\texttt{-extended}'' option to allow \texttt{CRUSH} to better preserve extended structures. We also use ``\texttt{-fixjumps}'' to remove glitches due to flux jumps in the SQUID amplifiers. The ``\texttt{-downsample}'' parameter is set to unity to prevent spatial averaging. We set the number of iterations using ``\texttt{-rounds=85}''. Additional information for the ``\texttt{-rounds}'' parameter is discussed in Appendix \ref{sec:conv}. {For each strip, all scans are analyzed together (8 scans for all fields but 1 for which only 4 scans were done). Instrumental polarization is then removed for each strip. The resulting image is shown in Figure~\ref{85rounds},} for which the individual strips have been co-added using the \texttt{CRUSH} \texttt{merge} algorithm. {For these data, the polarization angle is calculated by
\begin{equation}
    \phi=\frac{1}{2}\arctan\frac{U}{Q},
    \label{eq:phi}
\end{equation}
where the definitions of $Q$ and $U$ follow the IAU convention of (equatorial) North corresponding to an angle of 0$^\circ$, with $\phi$ increasing in the counterclockwise direction.}
The length of each pseudovector is proportional to the debiased fractional polarization, 
\begin{equation}
p=\sqrt{p_m^2-\sigma_p^2}.
\label{eq:p}
\end{equation}
\noindent  Here, $\sigma_p$ is the uncertainty in the measured fractional polarization, $p_m$ \citep{Serkowski1974}. In plotting Figure~\ref{85rounds} we apply cuts in polarization signal-to-noise, $p/\sigma_p>3$, polarization, $p<50\%$, and signal-to-noise of the total intensity (Stokes $I$), $I/\sigma_I>200$.\footnote{These constraints are consistent with the standard polarimetry cuts used by SOFIA.} These cuts yield 24,569 Nyquist-sampled detections of polarization.

\begin{figure*}
    \centering
    \includegraphics[height=0.5\textwidth]{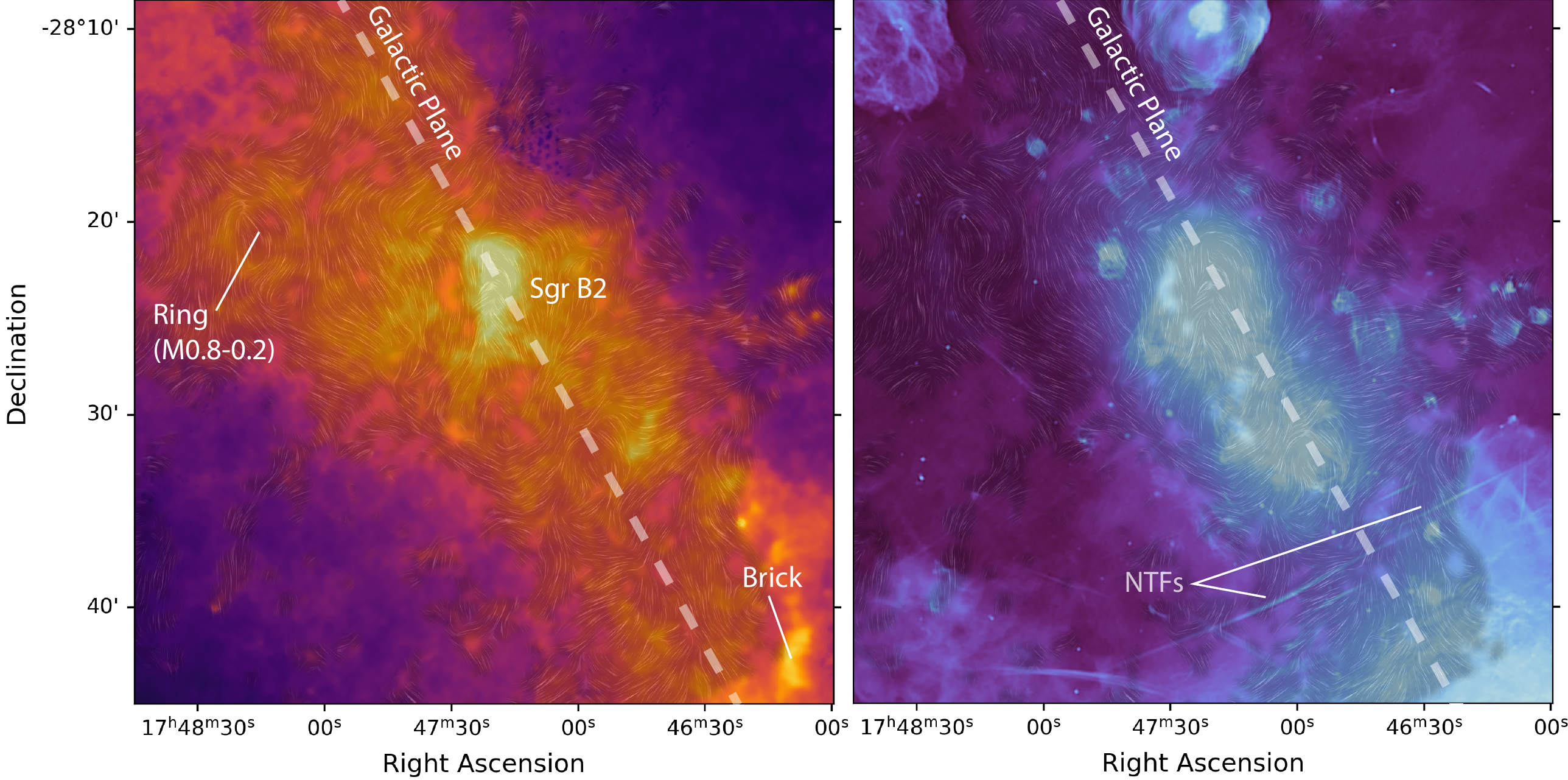}
    \caption{Line integral contour \citep[LIC;][see Figure \ref{85rounds-lic}]{Cabral1993}  representations of the magnetic field in the cool dust  superposed on two data sets. (Left) The dust emission from Herschel/SPIRE at 250 \micron\ is shown to highlight the region with high signal-to-noise \citep{molinari11}. (Right) The MeerKAT 20 cm (1 GHz) survey traces the non-thermal Synchrotron and thermal Bremsstrahlung emission \citep{heywood19}. Specifically, the long, straight NTFs are visible that trace magnetic fields in the intercloud medium of the Galactic center. Annotated in the Herschel 250 \micron\ image (Left) are the following sources: Sgr B2, the Brick, and the Ring (M0.8-0.2). The Ring source, covered in the DR1 release, will be discussed in detail in the FIREPLACE II publication (Butterfield et al. 2023, submitted). Identified in the MeerKAT 1 GHz image (Right) are a few NTF structures covered in the DR1 release. 
    }
    \label{fig:LICs}
\end{figure*}

Roughly 20\% of these detections (4,500 pseudovectors) are outside of the highly cross-linked central 30\arcmin\ of the map (black dashed box in Figure \ref{85rounds}) and are less reliable than those within the cross-linked area. These regions include the end of the scan parallel to the Galactic plane shown protruding to the upper left part of the map in Figure~\ref{85rounds} {(see Figure \ref{hitmap} for the cross-linking between scans)}.  We have found that an additional conservative cut in which we only consider vectors above a Stokes $I$ threshold, $I>5,000$ MJy sr$^{-1}$, provides a good conservative criterion for removing these pseudovectors. We briefly examine the statistics of the magnetic field pseudovectors in this lower intensity regime in Section \ref{sec:histogram} and apply this intensity cut in examining individual Galactic center clouds in Section~\ref{sec:sources}. 

{Polarized dust emission is a useful tool in probing interstellar magnetic fields. Ultimately, the interpretation of magnetic fields from the polarization relies on a complete understanding of the physics of the alignment process.  In most cases studied to date, the rotation axis of a grain is assumed to be preferentially aligned with the magnetic field direction as a result of radiative torques (RATs), and is usually assumed to be orthogonal to the plane-of-sky component of the magnetic field \cite[\eg,][]{andersson2015}.  This alignment mechanism is commonly referred to as ``$B$-RATs''. It has been posited that in some cases, the grain rotation axis may instead become aligned with the direction of the radiation vector ($k$-RATs) \citep[\eg,][]{Lazarian07, Tazaki_2017}.  There is ongoing work on the latter and ultimately, the FIREPLACE dataset may be useful for investigating $k$-RATs. However, for this paper, we will assume that the alignment is magnetic and for the remainder of the paper, we will show and interpret the inferred magnetic field direction, which is done by rotating the polarization pseudovectors by 90$^\circ$. Figure \ref{85rounds-crop} shows a larger version of the highly cross-linked central region from Figure \ref{85rounds}, in which the pseudovectors have been rotated to show the inferred magnetic field direction.}

\section{The GC Magnetosphere}
\label{Magnetosphere}


One of the key science goals of the FIREPLACE survey is to search for and characterize the connection between the field traced by the energetic particles within the non-thermal filaments and that in the dense molecular clouds. {The NTFs are mostly oriented perpendicular to the Galactic plane, suggesting a poloidal geometry for the Galactic center magnetic field.\citep[\eg,][]{LaRosa04}. However, the large-scale dust polarization studies by PILOT \citep{guan21} and ACTpol \citep{Mangilli19} show a field orientation that is more closely aligned with the Galactic Plane, suggesting a torodial configuration in the dust phase of the CMZ.}
The FIREPLACE survey DR1 dataset enables insight into {this inconsistency between the different observational studies.} 

The magnetic field directions that are inferred from the polarimetry data are plotted as line integral contours \citep[LICs;][]{Cabral1993} in Figure~\ref{fig:LICs} (a short discussion on LICs is included in Appendix \ref{sec:lic}). On the left, the magnetic field directions in the cool dust from the FIREPLACE survey are shown superposed on the Herschel/SPIRE 250 \micron\ map of the CMZ.  The Eastern \til0.5\degree\ Galactic region of the ``Twisted Ring'' structure, described by \citet{molinari11}, is seen in the lower right quadrant of the map.  The magnetic fields show no obvious indication of the coherence of this structure. In fact, in some places, the field seems to run perpendicular to the direction of these structures.

In the right panel of Figure~\ref{fig:LICs}, the FIREPLACE magnetic field measurements are superposed on the MeerKAT 20 cm emission \citep[1 GHz;][]{heywood19, heywood22}. Several NTFs are present in the field-of-view covered by our dataset. Towards the southwestern part of the map, the field in the dust seems to have a complex relationship with the NTF directions.  In some places, the field in the dust is parallel to the filaments; in others it is perpendicular.  We anticipate that the full FIREPLACE survey will enable statistical studies of the relationship between synchrotron features and dust polarimetry direction; however, this initial data set provides a hint of a possible relationship between the field in the dust within clouds and that in the intercloud medium. Additional data from the full FIREPLACE survey will increase the fidelity of these measurements near the edge of our current map with additional observation time and cross-linked observations. However, we can conduct a  preliminary quantitative test of a connection between the field in the dust and the field traced by the NTFs that is predominantly perpendicular to the Galactic plane.

Previous large-scale studies of the field in the cool dust at the Galactic center were limited to resolutions of 1\arcmin\ and coarser.  These maps have revealed that the field in the cool dust is (within $\sim$20\degree) parallel to the plane of the Galaxy. This was first reported by \cite{Novak2003} and more recently by surveys at 224 GHz with ACTpol \citep{guan21} and 240 \micron\ with PILOT \citep{Mangilli19}.  The observed parallel field motivated \cite{Novak2003} to postulate that the field in the cool dust may be a toroidal field that had been sheared from an initially poloidal field. In this picture, the NTFs trace the primordial poloidal field in regions where the rotational dynamical energy of the clouds is too weak to disrupt the field.

The larger, more recent surveys are sensitive to larger spatial scales \citep[e.g., PILOT;][]{Mangilli19}. As such, these surveys are potentially measuring the superposition of a considerable amount of polarized intensity along the line of sight towards the Galactic center, likely out to 1--2 kpc from the CMZ. Therefore, these field vectors may not be confined to the field within the central 100 pc.

Higher resolution measurements from the Kuiper Airborne Observatory \citep{Dotson00} and ground-based polarimetry \citep{Dotson10} have hinted that the field local to CMZ clouds may be more complicated and highly coupled to the dynamics of each particular region \citep{chuss03}; however these measurements had sufficient sensitivity for only the brightest cloud cores.

\begin{figure}
   \centering
   \includegraphics[width=0.48\textwidth]{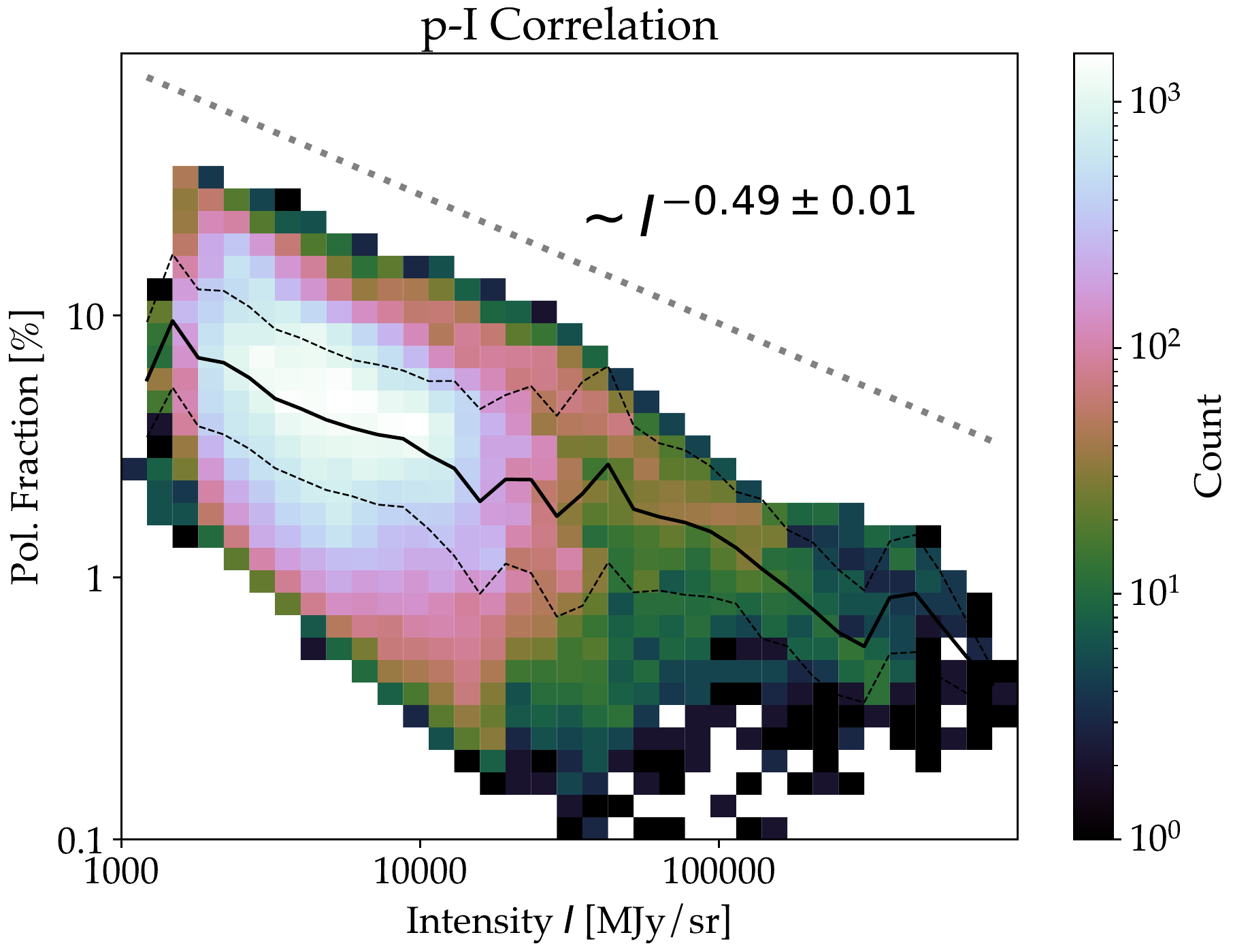}
   \caption{The 214 \micron\ fractional polarization plotted against the 214 \micron\ intensity for the 24,569 pseudovectors in the FIREPLACE DR1 survey from Figure \ref{85rounds}. Data points are presented as a log-log two-dimensional histogram with point density according to color scale. Black lines represent a smoothed version of the $p-I$ relation: for each bin value in $I$, median (solid) and standard deviation (dashed) values are calculated. The dashed grey line shows the slope of the best-fit line, $-0.49\pm0.01$, to the 214 \micron\ data points. 
   }
   \label{PvI}
\end{figure}

Figure \ref{PvI} shows the relationship between the 214 \micron\ intensity and the fractional polarization. There is a clear anti-correlation between the polarization and intensity, which can be fit with a slope -0.49$\pm$0.01 (dashed line in figure). \cite{chuss03} saw a similar distribution using CSO 350 \micron\ data toward the inner 50 pc of the CMZ which could be best fit with a slope of -0.67. This slight variation in the observed slope could be due to the observations being at different wavelengths or that our observations are more sensitive to lower fractional polarization vectors at lower intensities. Such anti-correlation has been seen in previous work \citep[\eg,][]{Fissel2016, Chuss2019} and can be due to either (or some combination of) loss of grain alignment in dense regions that are shielded from the interstellar radiation field (ISRF), magnetic field variation within the volume of the beam, or systematic effects due to sensitivity limitation and the nature of the Ricean distribution that describes the statistics of positive-definite quantities such as polarization \citep{Pattle19}.

As observed in Figure \ref{PvI}, most of the polarization data points have intensity values less than 20,000 \Mjsr\ (\til95\% of the entire sample). A majority of the data points above 20,000 \Mjsr\ are associated with the Sgr B2 complex. We do not detect any fractional polarization emission above 33\%, using the standard SOFIA/HAWC+ selection criteria.\footnote{Although there is a standard SOFIA/HAWC+ selection criterion that rejects any polarization vector having a fraction polarization above 50\%, this selection criterion is essentially moot, as our highest fractional polarization value, using the other selection criteria, is less than 33\% (Figure \ref{PvI}). However, we include this criterion in our list of selection cuts for consistency purposes.} 

\section{Bimodal Distribution of Magnetic Field Vectors}
\label{sec:histogram}

\begin{figure*}
   \centering
   \includegraphics[width=1.0\textwidth]{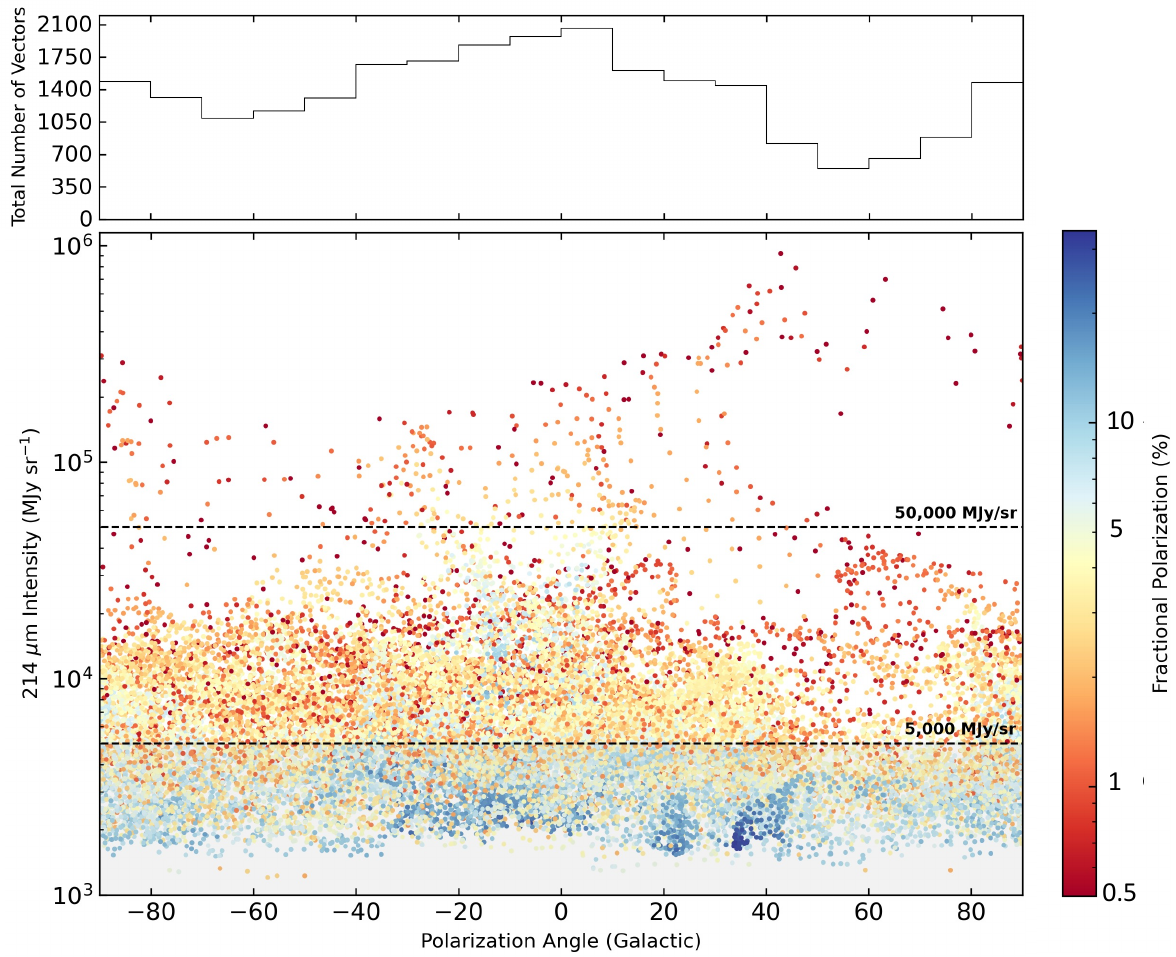}
   \caption{Distribution of the polarization angle, relative to the Galactic plane, for the 24,569 psudeovectors shown as a histogram (top) and as a function of the 214 \micron\ intensity (bottom). The colorbar is scaled logarithmically to illustrate the wide range of fractional polarization values, from 0.5\% to 33\% (maximum value in DR1). The histogram is binned by 10\degree\ increments in the polarization angle. 
   }
   \label{scatter-plot}
\end{figure*}

\begin{figure*}[p]
   \centering
   \includegraphics[width=0.9\textwidth]{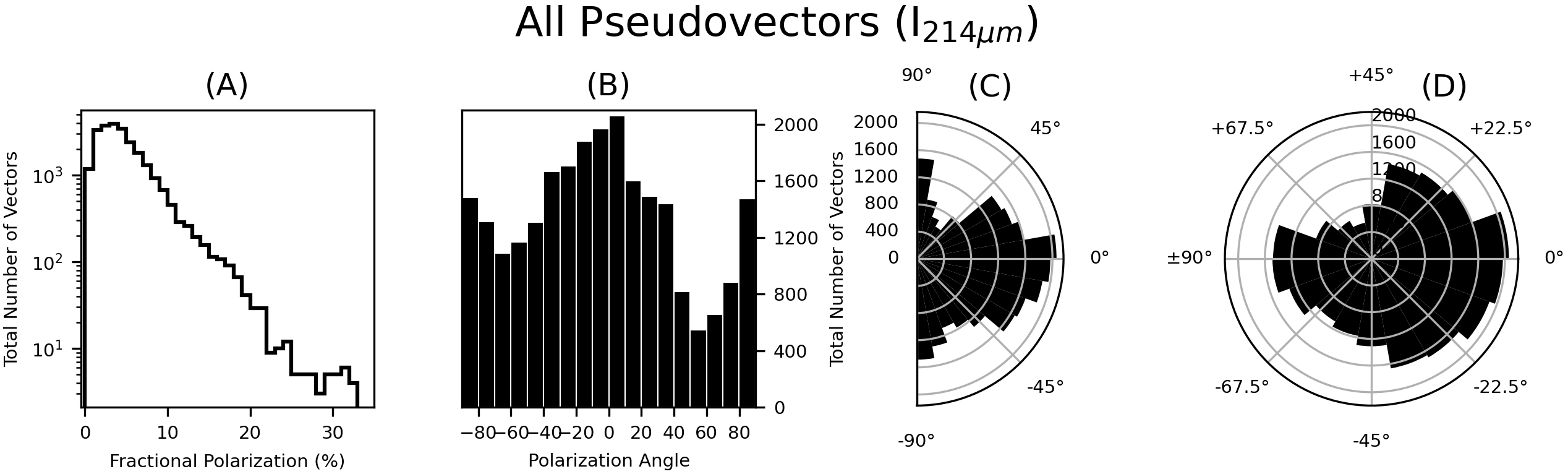}\\
   \includegraphics[width=0.9\textwidth]{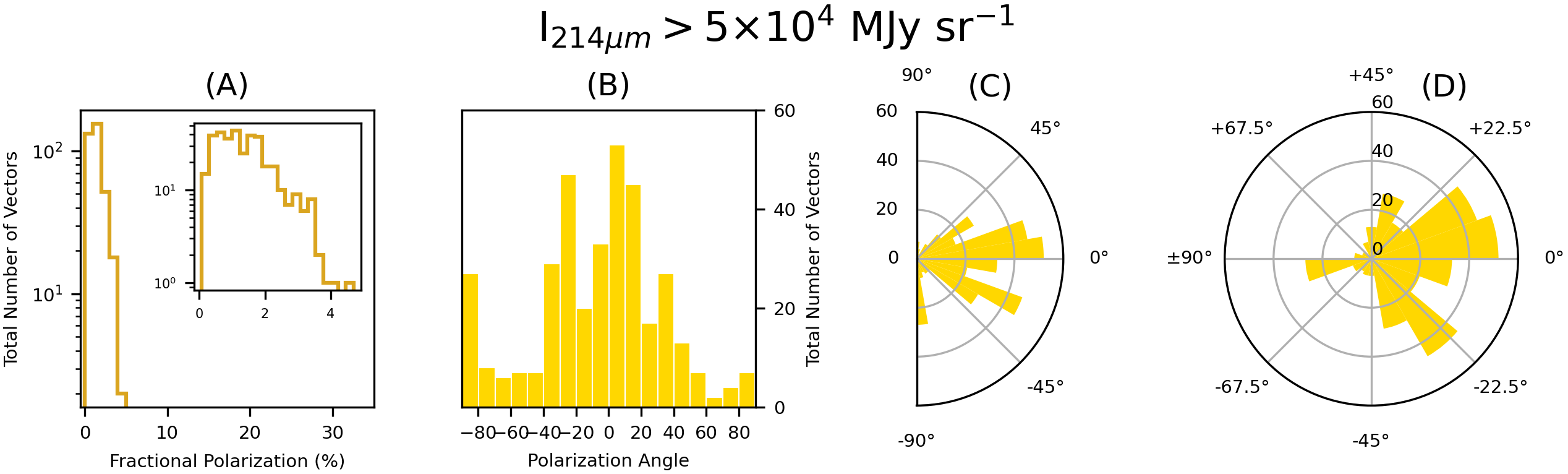}\\
   \includegraphics[width=0.9\textwidth]{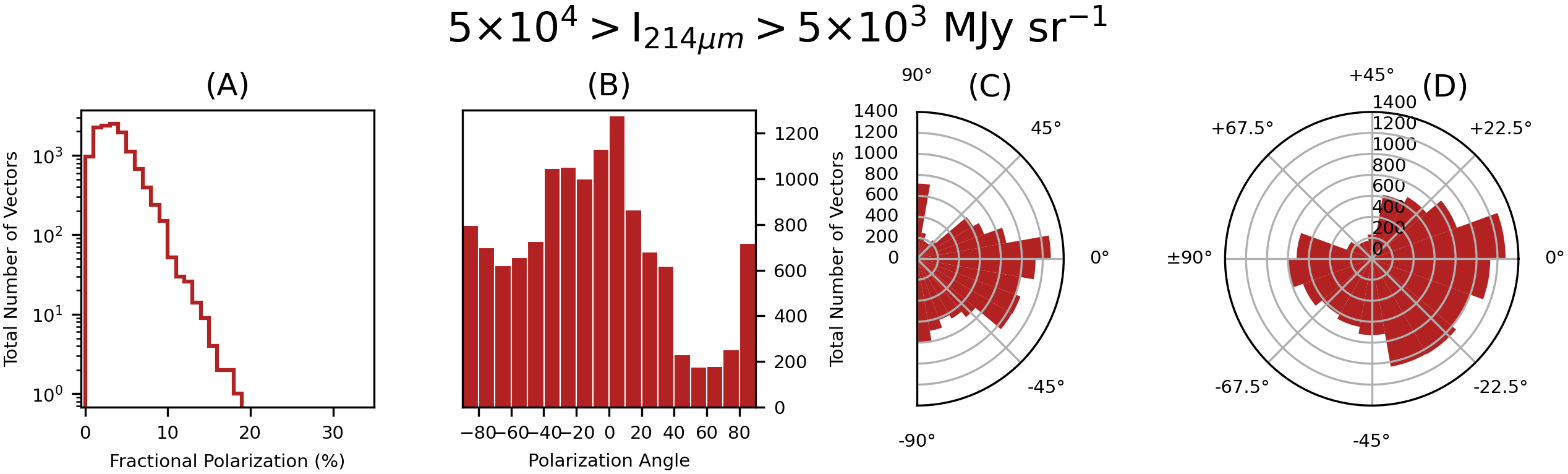}\\
   \includegraphics[width=0.9\textwidth]{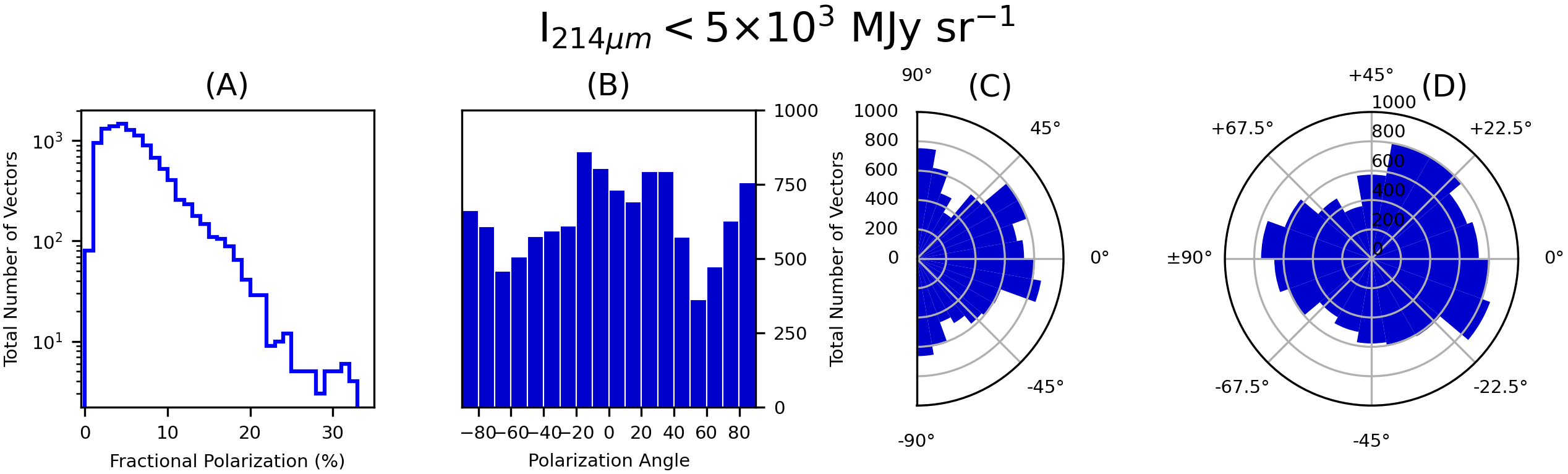}
   \caption{
   Distribution of the FIREPLACE polarization pseudovectors for {all pseudovectors} in DR1 (top row; black histograms; 24,569 pseudovectors) compared with the three different intensity bins, as titled: I$>$50,000 \Mjsr\ (yellow; 359 pseudovectors); 5,000$<$I$<$50,000 \Mjsr\ (red; 12,723 pseudovectors); I$<$5,000 \Mjsr\ (blue; 11,487 pseudovectors). These intensity ranges correspond with the dashed lines in Figure \ref{scatter-plot} (bottom). The columns show the follow information: (A) distribution of fractional polarization, summed over 1\% bins. (B) histograms showing the distribution of polarization angles in Galactic coordinates, similar to Figure \ref{scatter-plot} (top). A polarization angle of 0\degree\ corresponds to a magnetic field that is parallel to the Galactic Plane.  (C) Same distribution as Column B, but rotated to illustrate the plane-of-sky orientation. (D) Same distribution as Column C but ``wrapped" so that $\pm$90\degree\ is connected on the left side of the plot. 
   Here, pseudovectors on the left side of the plot ($\pm$90\degree) correspond to inferred magnetic fields that are oriented perpendicular to the Galactic Plane; those on the right hand side (0\degree) correspond to a field that is parallel to the Plane.
   The location of the pseudovectors for these three intensity ranges is shown in Figure \ref{hist-fig-map}.
   }
   \label{hist-fig}
\end{figure*}

The combination of resolution, sensitivity, spatial filtering (on scales comparable to the sizes of GC molecular clouds), {and coverage of the CMZ presented in our FIREPLACE survey is unprecedented in the literature}. {These SOFIA/HAWC+ observations of emission from cool dust enable a detailed measurement of the field structure over a broad region of the CMZ for the first time}. %
To quantitatively explore how the poloidal and toroidal fields in the cool dust component of the CMZ relate to each other, we plot the distribution of polarization vectors as a function of polarization angle relative to the Galactic Plane. This is analogous to the histogram of relative orientation \citep[HRO;][]{soler13} technique, but by comparing the magnetic field direction to the Galactic Plane rather than the local cloud structure.
Figure \ref{scatter-plot} (bottom) shows this plot for all pseudovectors that survived the initial cuts described in Section~\ref{obs} (i.e., standard polarimetry cuts used by the 24,569 Nyquist sampled SOFIA/HAWC+ pseudovectors). The data points are color-coded according to their fractional polarization value.
In this plot the polarization angle is measured counterclockwise from Galactic North (0\degree), conforming to the IAU polarization standard.\footnote{Note that this orientation is rotated from the equatorial coordinate system shown in the figures. To perform this coordinate transformation from equatorial to galactic for Figures \ref{scatter-plot} and \ref{hist-fig}, we added 63\degree\ to the observed polarization angle in the equatorial coordinate system measurement.} Since the magnetic field vector is oriented perpendicular to the polarization angle direction, a pseudovector with a polarization angle that is oriented towards Galactic North (0\degree) would have a magnetic field that is oriented in the East-West direction (i.e., parallel to the Galactic plane).
Due to the ``headless" nature of these pseudovectors, polarized emission that is oriented in the +90\degree\ direction is indistinguishable from a pseudovector oriented at -90\degree\ (i.e., pi-ambiguity). This results in the polarization angle, shown in Figure \ref{scatter-plot}, to essentially wrap around from +90\degree\ to -90\degree. We show this continuation of the polarization angle at $\pm$90\degree\ later in this section.

{As illustrated in Figure \ref{scatter-plot}, most of the higher fractional polarization values ($>$ 5\%, blue data points) are associated with lower intensity values ($<$5,000 \Mjsr). However, there are some higher fractional polarization values that extend upwards to around 50,000 \Mjsr. Most of these higher fractional polarization values that are above 5,000 \Mjsr\ have polarization angles  from roughly -35\degree\ to +10\degree. There is also a small fraction of these higher fractional polarization values that have polarization angles around $\pm$90\degree. 
These higher fractional polarization values could be an indication that the field is relatively well aligned at those locations or may be closely associated with locations experiencing shearing, similar to effects observed in \cite{Guerra23}. We will discuss the sources associated with these higher fractional polarization vectors in Section \ref{sec:sources}.}

{Above 50,000 \Mjsr, the fractional polarization values are generally lower, with values less than 5\%. Below 5,000 \Mjsr, the fractional polarization values are above 1\%, however this could be due to a selection effect of these low values not meeting the SOFIA standard cutoffs.}

{Figure \ref{scatter-plot} (top) shows the sum of all the data points from the bottom scatter plot, binned into 10\degree\ histogram bins. As shown here, there is a bimodal distribution, with most of the vectors being either parallel or perpendicular to the Galactic plane (with a somewhat higher population being parallel). To quantitatively investigate this bimodality, we break up the data into three intensity bins (I$<$5,000 \Mjsr, 5,000$<$I$<$50,000 \Mjsr, I$>$50,000 \Mjsr; dashed lines in Figure \ref{scatter-plot}, bottom) to investigate this bimodality further.}

\begin{figure*}[t!]
\centering
\includegraphics[width=0.95\textwidth]{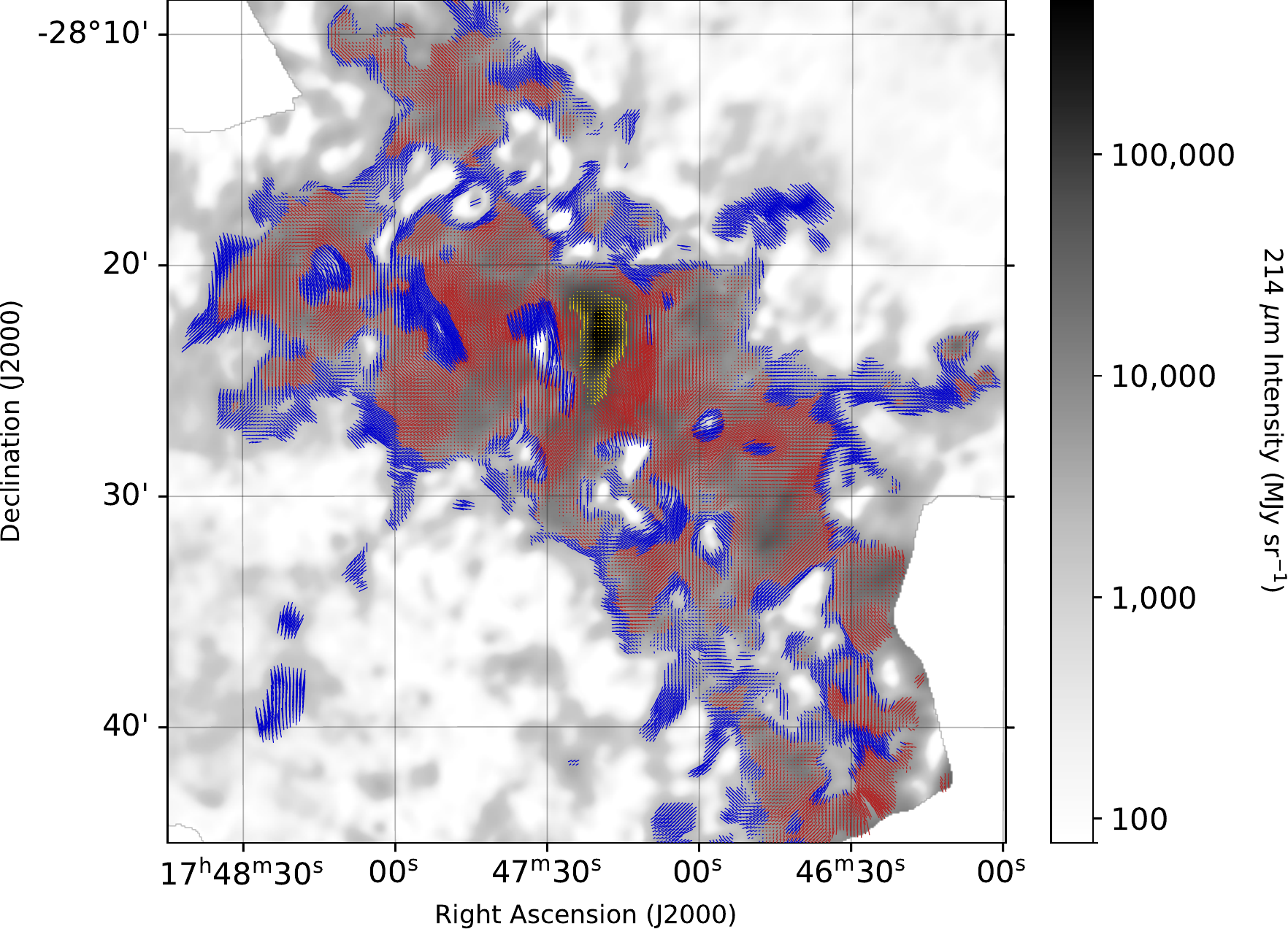}
\caption{Distribution of the polarization pseudovectors in DR1, colorized by the three intensity bin ranges shown in Figure \ref{hist-fig}: I$<$5,000 \Mjsr\ (blue); 5,000$<$I$<$50,000 \Mjsr\ (red); I$>$50,000 \Mjsr\ (yellow). This field-of-view is identical to Figures \ref{85rounds-crop}, \ref{fig:LICs}, \ref{pol-fig} and \ref{85rounds-lic}.  
}
\label{hist-fig-map}
\end{figure*}

Figure~\ref{hist-fig} shows a comparison of these three intensity bins (bottom three rows; titled by their intensity bin) to the total distribution for all pseudovectors in FIREPLACE DR1 (top row).
{The spatial location of the pseudovectors associated with these different intensity bins (yellow, red, and blue) is shown in Figure \ref{hist-fig-map}.}
{Column (A) {in Figure \ref{hist-fig}} shows the distribution of the fractional polarization from 0 to 35\%. As shown in Figure \ref{PvI}, there are no pseudovectors above 33\% that meet the SOFIA standard cuts. Column (B) shows a histogram of relative orientation (HRO) relative to the Galactic Plane. The top figure in this column is identical to Figure \ref{scatter-plot}, top. Column (C) shows the same distribution as Column B, but with a curved polarization angle axis to show the plane-of-sky orientation. Column (D) shows the same distribution as Column C but ``wrapped" so that $\pm$90\degree\ is connected on the left side of the plot. } 

{As shown in Column A, we observe a steeper slope in the fractional polarization for the higher intensity regimes (e.g., $>$50,000 \Mjsr) compared to the lower intensity regimes (e.g., $<$5,000 \Mjsr). This falloff in the fractional polarization could be the result of depolarization towards denser sources. We discuss potential mechanisms for this depolarization in more detail in Section \ref{depol}. In all three intensity bins, the majority of vectors have fractional polarization values around 2\%.}
{Additionally, the bimodal distribution, observed in Figure \ref{scatter-plot} (top), is also detected in all three intensity bins (see Columns B, C and D in Figure \ref{hist-fig}). In the two highest intensity bins, emission above 5,000 \Mjsr, the number of pseudovectors in a single 10\degree\ bin peaks around 0\degree\ (\ie, parallel to the Galactic Plane). Furthermore, the highest intensity bin ($>$50,000 \Mjsr) is almost exclusively weighted in the direction more consistent with the plane (See Column D), with only 20 per cent of vectors being more closely associated with the perpendicular field direction. 
}

There also appears to be a slight excess of vectors towards the left side of the 0\degree\ peak in the histogram plots - around polarization angles of -20\degree\ (see Figure \ref{hist-fig}, Column B). As shown in column C, this is oriented in a direction similar to that of the large-scale field component observed in the PILOT \citep{Mangilli19} and ACTpol \citep{guan21} surveys.\footnote{Although the PILOT survey cites a +22\degree\ orientation angle, it conforms to the COSMO convention typically used in the \textit{Planck} collaboration. The COSMO convention uses the HEALpix software package which defines the polarization angle as increasing clockwise \citep{gorski05}.} 
Additional analysis using the complete FIREPLACE survey (DR2) is needed to confirm whether this excess in the number of pseudovectors around -20\degree\ is observed across the entire CMZ.

\begin{figure*}[t]
\centering
\includegraphics[width=1.0\textwidth]{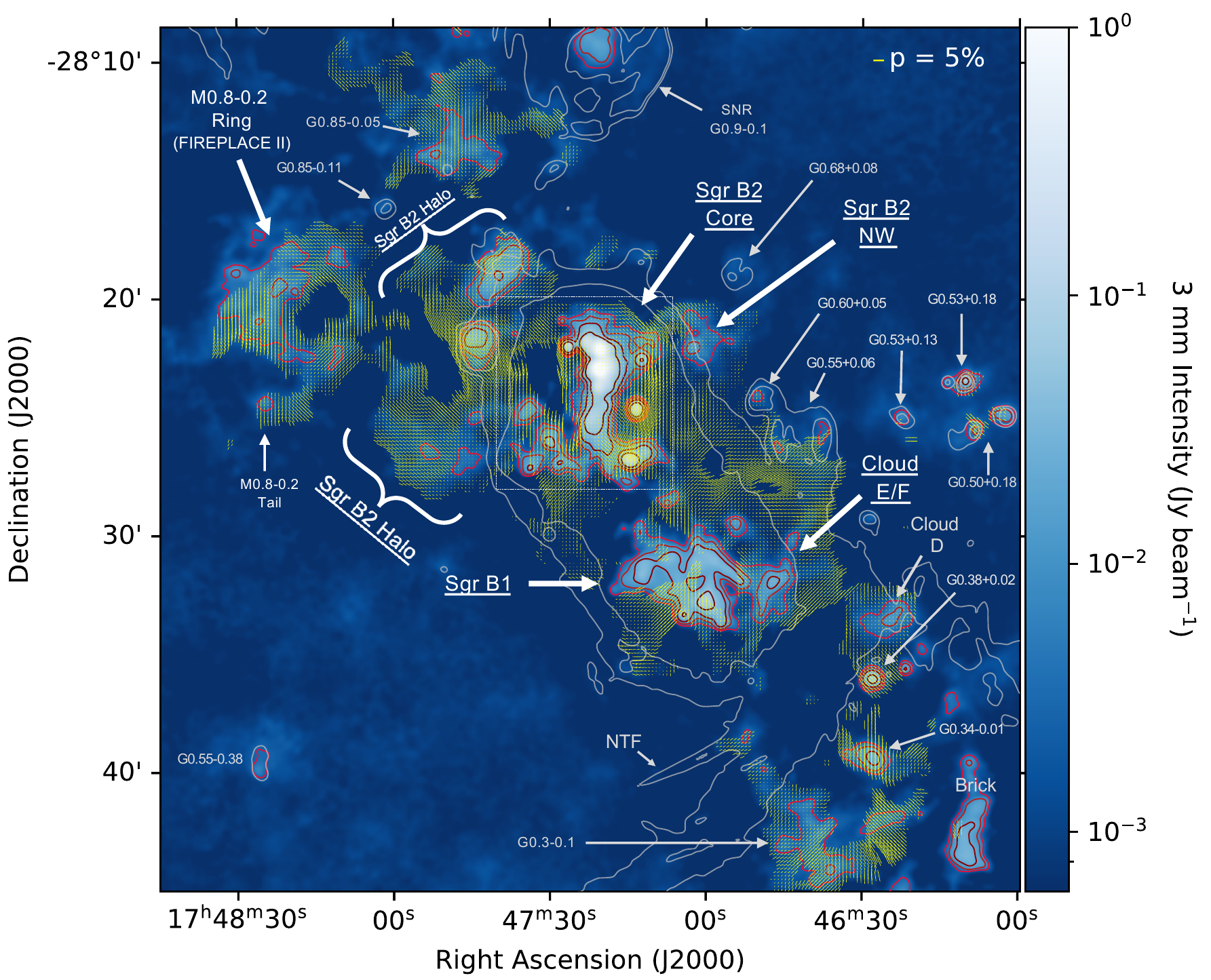}
\caption{SOFIA 214 \micron\ magnetic field directions from the FIREPLACE survey (same field-of-view as shown in Figures \ref{hitmap} (blue box), \ref{85rounds} (black box), \ref{fig:LICs}, and \ref{hist-fig}), for emission above 5,000 \Mjsr\ (\eg, red and yellow pseudovectors in Figure \ref{hist-fig-map}), overlaid on the GBT+MUSTANG 3 mm (90 GHz) microwave emission \citep[MUSTANG Galactic Plane Survey (MGPS);][]{ginsburg20}, smoothed to an angular resolution of 18\arcsec. Red contours show the MGPS data at 10$\sigma$, 20$\sigma$, 40$\sigma$, 100$\sigma$, and 400$\sigma$ (for an rms value of 0.5 m\jybe).  Grey contours show the MeerKAT 1 GHz data at 40\% and 50\% of the peak emission \citep[][smoothed to the 18\arcsec\ resolution of the FIREPLACE survey]{heywood19}. Annotated are several well-known sources in the Galactic Center. The underlined sources are discussed in detail in Section \ref{sec:sources}. Sources located in and around the SgrB2 Core (white box) are annotated on Figure \ref{Sgb-b22}.  
}
\label{pol-fig}
\end{figure*}

In higher intensity regions, which corresponds to dense cloud regions, the field tends to be much more likely parallel to the Plane. This connection between denser cloud regions and orientation of the field parallel to the plane supports the conclusion of \citet{chuss03}. They hypothesized that the fields in denser regions are dragged into a roughly toroidal configuration by the shearing of the clouds by the orbital motion, so that the projected magnetic field vectors tend to be parallel to the plane. 

In lower intensity regions (I$<$5,000 \Mjsr; blue pseudovectors), which correlate to lower-density regions, there is a more uniform distribution of magnetic field directions (Figure \ref{hist-fig}, Column B, C and D) than the other two intensity bins. Furthermore, the fractional polarization plots for this lower intensity regime (Column A in Figure \ref{hist-fig}), show fractional polarization values above 20\% - typically the highest fractional polarization values observed in ISM clouds. Due to the large sample size of the FIREPLACE DR1 Survey (24,569 pseudovectors), the detection of 129 pseudovectors above 20\% (roughly 0.5\% of the entire sample), may be possible. Additional probabilistic analysis is needed to determine if these higher fractional polarization values ($>$20\%) are significant. Since these regions are within the sensitivity limit of the FIREPLACE survey, we anticipate that with the addition of more data and additional careful testing of the pipeline we can study this emission more thoroughly in future work. Therefore, we implement a conservative data cut limit on total intensity ($I>5,000$ \Mjsr) for DR1 that we adopt in Section \ref{obs} and apply in Section \ref{sec:sources}.

\begin{figure*}
\centering
\includegraphics[width=1.0\textwidth]{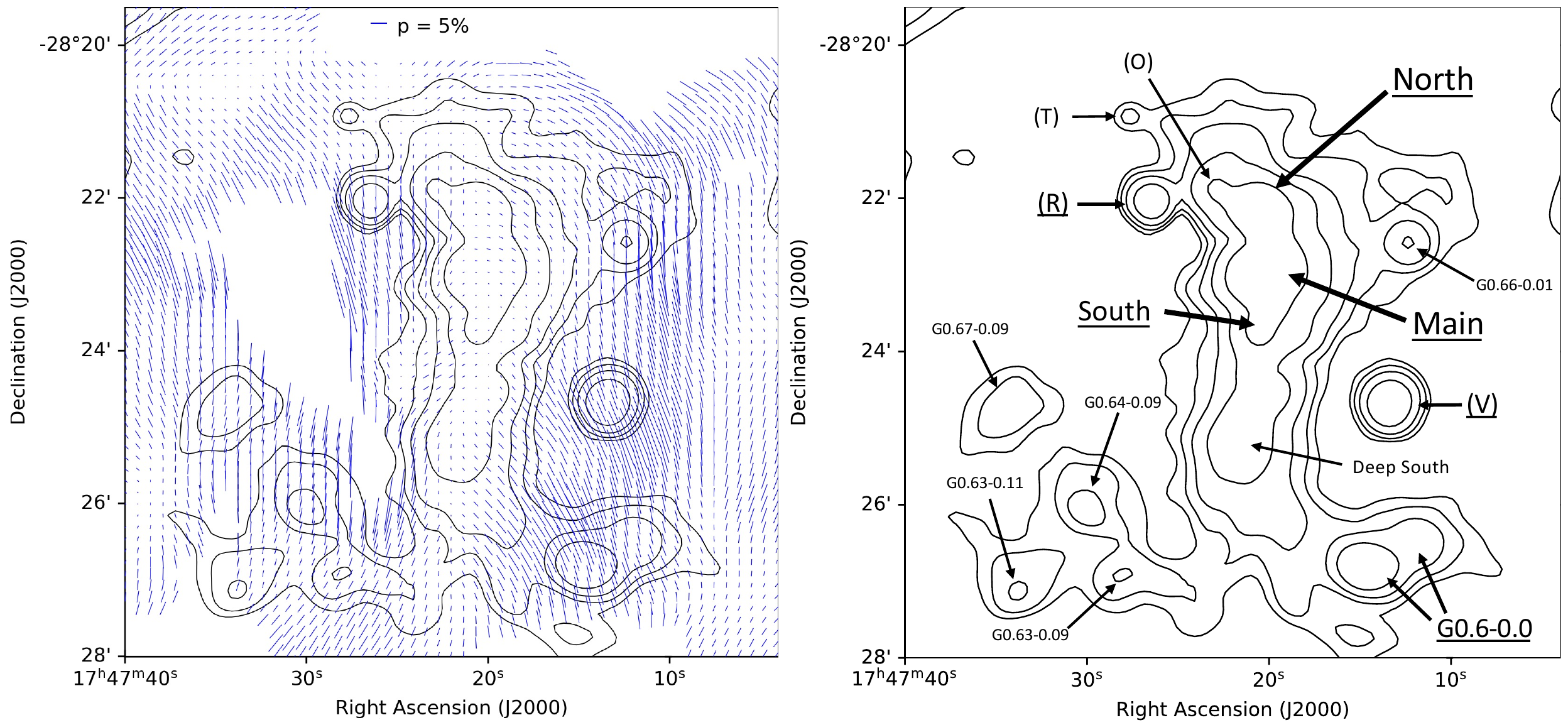}
\caption{(left) 214 \micron\ magnetic field directions (blue pseudovectors; for 214 \micron\ emission above 5,000 \Mjsr) in the SgrB2 Core (white box from Figure \ref{pol-fig}). Black contours show the MGPS data \citep[][smoothed to an 18\arcsec\ resolution]{ginsburg20} at 10$\sigma$, 20$\sigma$, 40$\sigma$, 100$\sigma$, and 400$\sigma$ (for an rms value of 0.5 m\jybe). (right) same contour levels as shown at left with annotations of compact sources in this field of view. Several of these underlined features are discussed in Section \ref{sec:sgrb2-core}. 
}
\label{Sgb-b22}
\end{figure*}

\section{Magnetic Field Structure in Notable CMZ Clouds}
\label{sec:sources}

The high angular resolution and high sensitivity of the FIREPLACE survey, compared with previous polarization surveys of the CMZ, enables us to investigate the magnetic field geometry  in several prominent regions of the CMZ.  Figure \ref{pol-fig} {shows a comparison of the 214 \micron\ magnetic field vectors with the MeerKAT 1 GHz and MUSTANG 3 mm datasets and} highlights several prominent CMZ clouds included in the DR1 dataset. {Annotated sources that are underlined (e.g., Sgr B2 Core, Sgr B1, Cloud E/F, etc) are sources} that we discuss in the following sections. 
Due to the unique nature of the \ring\ ring, this source will be discussed separately in a forthcoming paper (FIREPLACE II, Butterfield et al. 2023, submitted).

\subsection{Sgr B2 Core}
\label{sec:sgrb2-core}

The Sgr B2 Core is one of the most intense star forming regions in the Milky Way Galaxy, with a star formation rate of \til0.028–0.039 \msuny\ \citep[\eg,][]{Belloche13}, qualifying Sgr B2 as a mini-starburst. The Sgr B2 Core is the brightest region in the FIREPLACE survey, with 214 \micron\ total intensity values exceeding 100,000 \Mjsr. As mentioned in the context of data validation (Section~\ref{sec:sb2}), the polarization toward the Sgr B2 Core and complex is quite consistent with that at 350 \micron\ (see Figure \ref{fig:Sgrb2} for this comparison). Figure \ref{Sgb-b22} shows an enlarged view of the Sgr B2 Core, identified in Figure \ref{pol-fig}, with annotations of several sources that we will discuss in the following sections. 

\begin{figure*}
\centering
\includegraphics[width=1.0\textwidth]{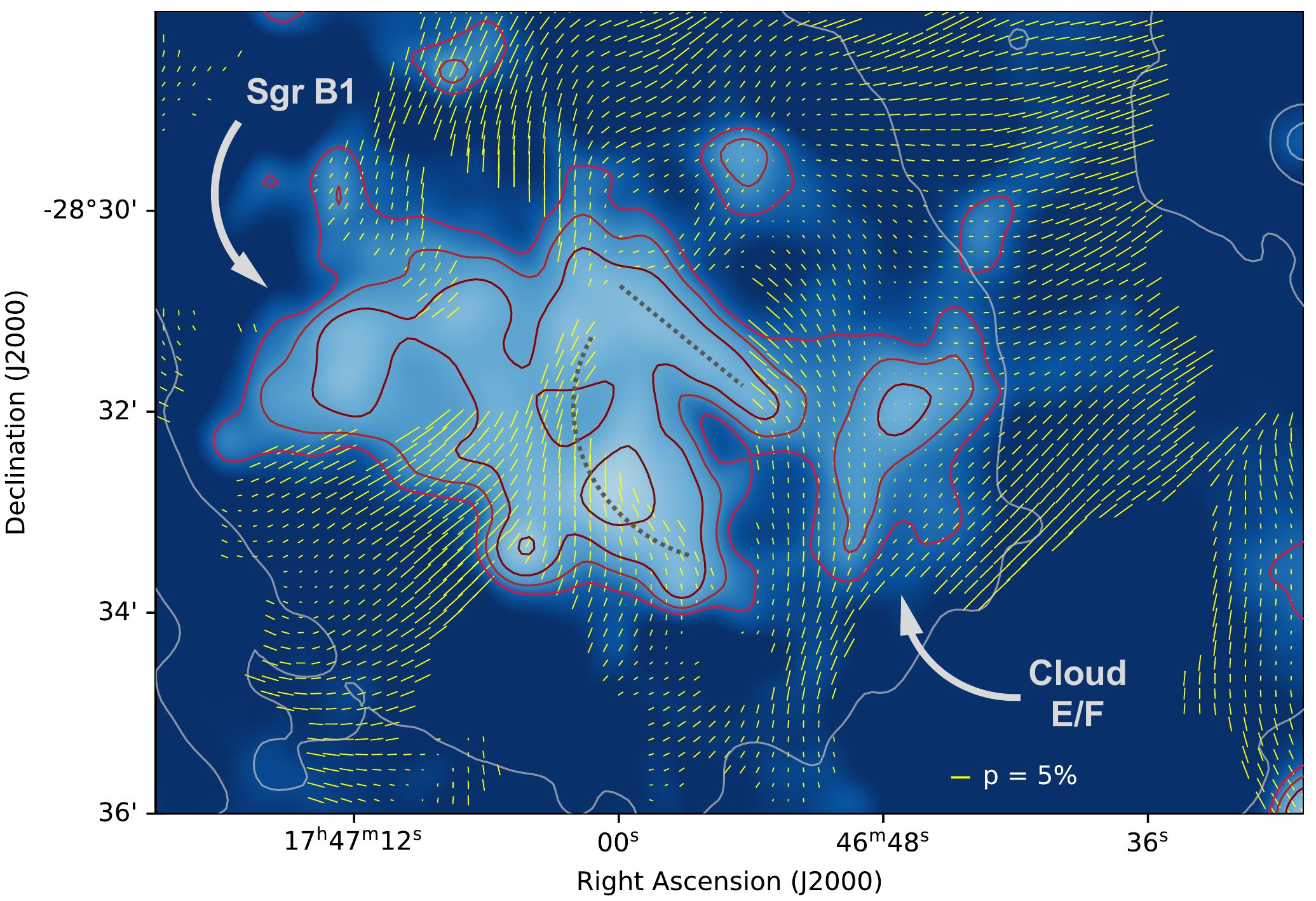}
\caption{214 \micron\ magnetic field directions (yellow pseudovectors; for 214 \micron\ emission above 5,000 \Mjsr) in the Sgr B1 and Dust Ridge Cloud E/F, overlaid on the 3-mm MGPS emission \citep[][smoothed to an angular resolution of 18\arcsec]{ginsburg20}. Red contours show the MGPS data at 10$\sigma$, 20$\sigma$, 40$\sigma$, and 100$\sigma$ (for an rms value of 0.5 m\jybe).  Grey contours show the MeerKAT 1 GHz data at 40\% and 50\% the peak emission \citep[][smoothed to the 18\arcsec\ resolution of the FIREPLACE survey]{heywood19}. The curved and straight dashed lines in the Sgr B1 complex illustrate the Ionized Rim and Ionized Bar, respectively \citep{mehringer92}. 
}
\label{sgrb1ef}
\end{figure*}

\subsubsection{Sgr B2: North, Main and South}
\label{depol}

The main core of Sgr B2 contains three main sources: North (G0.680-0.028), Main (G0.668-0.036), and South (G0.659-0.042); as identified in Figure \ref{Sgb-b22} (right). This region of the Sgr B2 complex contains the hot, young molecular cores that are actively forming stars \citep[\eg,][]{Belloche13,ginsburg18}. The fractional polarization towards these sources in the Sgr B2 Core is lower ($<$1\%) when compared to the periphery of the cloud (\til1$-$5\%; Figure \ref{Sgb-b22}, left). These inner regions of the clouds contain higher column densities than the peripheries, with values reaching 10$^{24}$ cm$^{-2}$ (see \cite{molinari11}, their Figure 4, and \cite{Marsh2015} for a column density map using Herschel data).\footnote{Maps of the column density from \cite{Marsh2015} can be found here: \hyperlink{http://www.astro.cardiff.ac.uk/research/ViaLactea/}{http://www.astro.cardiff.ac.uk/research/ViaLactea/}. } This decrease in the fractional polarization towards the interiors of these dense clouds is likely due to some combination of three effects. First, in the interior of these clouds the grain alignment may be less efficient.  Loss of grain alignment has been observed in the cores of clouds \citep{Santos2019}. This has been explained in the context of radiative torque (RAT) theory via shielding from the interstellar radiation field. On the other hand, \cite{Chuss2019} have found that in OMC-1, it is not necessary to invoke loss of grain alignment to explain depolarization in all clouds. \cite{Michail2021} suggest that this may be due to embedded sources supplying the optical/UV light to spin up the grains. Sgr B2 contains numerous embedded sources \citep[\eg][]{ginsburg18} and may be similar to OMC-1 in this respect.

The second factor is that sight lines towards the Sgr B2 complex may include a superposition of multiple field directions in different portions of the cloud along the line of sight. It is the only known source to exhibit polarization by absorption at far-infrared wavelengths \citep[][who explains its polarization spectrum as a result of such a superposition]{Dowell1997}. The high density in the cores of Sgr B2, combined with differing magnetic fields in the cores and the cooler surrounding envelope, lead to a complicated wavelength-dependent polarization.  Multi-wavelength polarimetry at shorter wavelengths would provide a good tool for investigating both of these possibilities.

The final mechanism that could be contributing to the depolarization near the center of the cloud is that the field structure may be tangled below the resolution scale of the HAWC+ beam.  Higher resolution observations such as those from ALMA can be used to unambiguously determine the sub-beam magnetic field structure.

\subsubsection{Sgr B2: \hii\ Regions (R, V, \& G0.6)}

In the Sgr B2 field, there are also two bright, compact 3 mm sources `V' (G0.631-0.028) and `R' (G0.693-0.046) are coincident with relatively high fractional 214 \micron\ polarization, with typical values around 10\%, and some pseudovectors reaching values as high as 15\%. These sources `V' and `R' are bright, compact \hii\ regions, with high ionization rates (10$^{48-49}$ photons s$^{-1}$) and are suggested to be associated with O-type stars \citep[see Table 2 in][]{mehringer93}.

The southern G0.6-0.0 source, comprised of four ultracompact \hii\ regions \citep[$<$3\arcsec, $<$0.1 pc;][]{mehringer92}, also shows relatively high fractional polarization at 214 \micron, with values around 10\% and, in a few pixels, reaching up to 15\%. These \hii\ regions are bright at radio frequencies (\til5 GHz) and are rich with radio recombination line emission \citep{mehringer92, myT}. \cite{mehringer92} also measured a high ionization rate of 10$^{48-49}$ photons s$^{-1}$ for each of the four ultracompact \hii\ regions, indicating they are also associated with O-type stars (see their Table 4).

The pseudovectors spatially aligned with the G0.6-0.0 ultracompact \hii\ regions and the V and R sources near the Sgr B2 complex are associated with the high fractional polarization vectors discussed in Section \ref{sec:histogram} and displayed in Figure \ref{scatter-plot}. These pseudovectors are generally aligned with the Galactic Plane and are relatively uniform.
The sources themselves contain little dust, as evidenced by the lack of detection at far-infrared wavelengths (see Figure~\ref{fig:LICs}, left). This, in addition to the fact that the inferred magnetic field geometry shows little evidence of correlation with the morphology of the compact sources, indicates that the magnetic fields probed by the polarization measurements are located in the diffuse dust surrounding each of these \hii\ regions.

However, the local enhancement of the fractional polarization in the vicinity of each of these sources may be an indication that the UV-optical radiation from the stars in these regions is enhancing the grain alignment efficiency in the surrounding dust. This is consistent with Radiative Torques (RAT) alignment theory \citep{andersson2015} that posits that UV and optical radiation from stars drives the process by which grains become magnetically aligned via the transfer of angular momentum to the grains from the radiation field.

\subsection{Sgr B1 and Dust Ridge Cloud E/F}
\label{sec:sgrb1ef}

Figure \ref{sgrb1ef} shows the dust polarization around the SgrB1 complex, compared to the MeerKAT 1 GHz emission (grey contours) and the GBT/MUSTANG 3 mm emission (red contours). Note that the Sgr B1 \hii\ region (G0.5-0.05) is spatially offset from the Dust Cloud E/F (M0.47-0.01; \eg, red contours in Figure \ref{sgrb1ef}). The observed close proximity of the Sgr B1 \hii\ region and the Dust Ridge Cloud E/F (\til1\arcmin) may be the result of a projection effect. This projection effect could be caused by an overlap along our line of sight by large-scale orbital stream \citep[see Figure 4 in][for this orientaion]{barnes17}. Thus, Sgr B1 and Cloud E/F could be unique sources, and therefore we discuss them independently.

\subsubsection{Sgr B1}

The Sgr B1 \hii\ region (G0.5-0.05) has been studied in great detail at radio wavelengths, using the VLA \citep[\eg][]{mehringer92,myT}, and infrared wavelengths, using SOFIA FIFI-LS, FORCAST, and upGREAT observations \citep[\eg][]{simpson18,simpson21, Hankins20, Harris21}. While the magnetic field polarization vectors do not appear to be associated with the \hii\ region, they appear to trace the `Ionized Rim' (curved dashed line in Figure \ref{sgrb1ef}). The fractional polarization values in the cloud are  \til1--5\%. \citet{mehringer92} argue that the \hii\ region is the result of O-type stellar radiation, with ionization rates of 10$^{49}$ photons s$^{-1}$ (see their Table 4). They further discuss Sgr B1 as being an evolved \hii\ region with an age of \til10$^6$ yr.

Unlike the compact \hii\ regions observed near Sgr B2 (\eg, V and R), Sgr B1 doesn't appear to show high fractional polarization - having typical values $\leq$5\%. This could be due to the fact that Sgr B1 is an evolved \hii\ region and may have evacuated the surrounding dust that would result in polarized emission at 214 \micron. This would imply that the high fractional polarization emission observed in the V and R \hii\ regions near Sgr B2 may be a relatively short evolutionary stage of the luminous stellar objects that contribute to grain alignment, thereby producing the high polarization.

There is an additional compact source in our DR1 coverage (G0.38+0.2; Figure \ref{pol-fig}) that shows high fractional polarization, however, this source is located near the edge of our coverage and contains only a single mapping scan (Figure \ref{hitmap}). Analysis of this source, and potentially other compact \hii\ sources in the complete FIREPLACE survey could give insight on the nature of the highly polarized emission (DR2; Par{\'e} et al., in prep). 

\subsubsection{Dust Ridge Cloud E/F}

Cloud E/F (M0.47-0.01) is part of a larger CMZ structure known as the Dust Ridge \citep{longmore13,Kru15}, which contains other CMZ clouds such as the Brick and Sgr B2 (see Figure \ref{3color} and Figure \ref{fig:LICs}, left). Cloud E/F has been argued to be in a younger evolutionary stage compared to the Sgr B2 Core \citep[e.g.,][]{longmore13,Kru15}. \cite{barnes19} observed Cloud E/F at high angular resolution using ALMA and detected numerous compact cores within the interior of the cloud, indicating star formation could be underway within the densest regions.

Similar to the Sgr B2 Core (Figure \ref{Sgb-b22}), the E/F cloud shows lower fractional polarization ($<$1\%) towards the bright center of the cloud and higher fractional polarization (\til1--5\%) towards the periphery (Figure \ref{sgrb1ef}). Therefore, Cloud E/F could be experiencing depolarization conditions similar to those in Sgr B2 (see Section \ref{depol} for this discussion).

\begin{figure}
\centering
\includegraphics[width=0.45\textwidth]{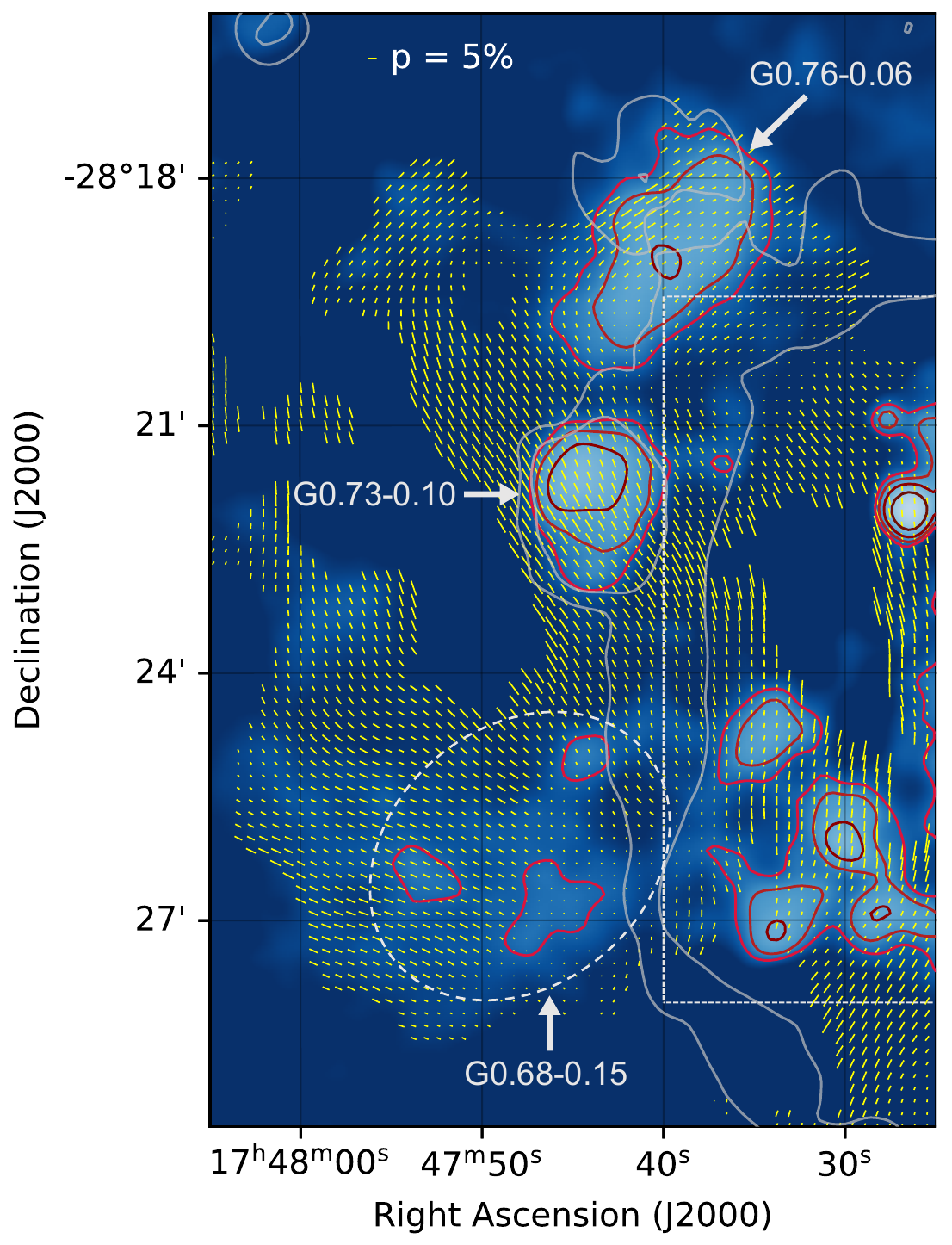}
\caption{SgrB2 Halo region: yellow pseudovectors show magnetic field directions for 214 \micron\ surface brightness above 5,000 \Mjsr, overlaid on the 3 mm MGPS emission \citep[][smoothed to an angular resolution of 18\arcsec]{ginsburg20}. Red contours show the MGPS data at 10$\sigma$, 20$\sigma$, 40$\sigma$, and 100$\sigma$ (for an rms value of 0.5 m\jybe). Grey contours show the MeerKAT 1 GHz data at 40\% and 50\% the peak emission \citep[][smoothed to the 18\arcsec\ resolution of the FIREPLACE survey]{heywood19}. Annotated are the 3 sources discussed in Section \ref{sec:sgrb2-halo}: G0.68-0.15 (white oval), G0.73-0.10, G0.76-0.06. White box shows the cross-over coverage of the Sgr B2 Core (Figure \ref{Sgb-b22}).
}
\label{Sgb-halo}
\end{figure}

\subsection{Sgr B2 Halo}
\label{sec:sgrb2-halo}

The Sgr B2 Halo is an extended structure (12\arcmin\ in length; \til30 pc), located east of the Sgr B2 Core \citep[see Figure \ref{pol-fig} in this manuscript and discussion in][]{mills17}. Figure \ref{Sgb-halo} shows a cropped version of Figure \ref{pol-fig}, with a focus on the Sgr B2 Halo, and annotated sources within the structure. This region has a `S'-like morphology in the 214 \micron\ emission, where the Southern half of the cloud curves eastward (Figure \ref{Sgb-halo}).  The Northern region of the cloud bends at roughly a 45\degree\ angle, creating a somewhat `S'-like shape in the cloud morphology (Figure \ref{Sgb-halo}). The magnetic field generally follows this large-scale `S'-like structure of the cloud. In general, the Sgr B2 Halo cloud has fractional polarization values 1--5\%. However, in the middle region of this cloud, around the 3 mm emission source G0.73-0.10, the fractional polarization increases to \til8\%. This region corresponds to the middle of the `S'-like structure of the cloud.

The 3 mm emission shows a very different character in the Northern portion of the cloud compared to the Southern region. The Northern region of the cloud contains bright 3 mm emission (G0.76-0.06; Figure \ref{Sgb-halo}), whereas the Southern region of the Sgr B2 Halo contains dense, cold clumps of gas and dust (G0.68-15; Figure \ref{Sgb-halo}). The region around G0.68-0.15 (white oval in Figure \ref{Sgb-halo}) has low fractional polarization, with the largest 3 mm clump showing some of the lowest fractional polarization values in the cloud ($<$1\%). At the middle of the cloud \citep[G0.73-0.10;][]{LaRosa00}, the polarized emission is strongest (up to 10\% in some locations). The G0.73-0.10 source is shown to have numerous compact 1 mm sources ($<$10\arcsec; \til20 sources) compared with G0.76-0.06, which has 4, and G0.68-15, which has 0 \citep[CMZoom Survey II, see Figure 19 in][]{Hatchfield20}. Furthermore, radio recombination line emission was detected from the G0.73-0.10 source by \citet{PM75}, indicating that it is thermal in nature and likely an \hii\ region (see their Figure 2). The high fractional polarization values detected towards this source could be the result of the luminosity of the star producing the \hii\ region providing radiative alignment, similar to the high fractional polarization detected towards Sgr B1, G0.6-0.0, and the Sgr B2 sources V and R. The G0.73-0.10 source is slightly more extended than G0.6-0.0 and the Sgr B2 sources V and R, but not as extended as the Sgr B1 complex. Additionally, the fractional polarization emission from G0.73-0.10 (\til8\%) is higher than values observed towards Sgr B1 (\til5\%), but not as high as the fractional polarization values measured in G0.6-0.0, and the Sgr B2 sources V and R (10-15\%). This observation could be an indication that the fractional polarization values are inversely related to the extent of the sources where the radiative grain alignment is lessened in the larger sources due to the radiation field being diluted.

\begin{figure}
\centering
\includegraphics[width=0.45\textwidth]{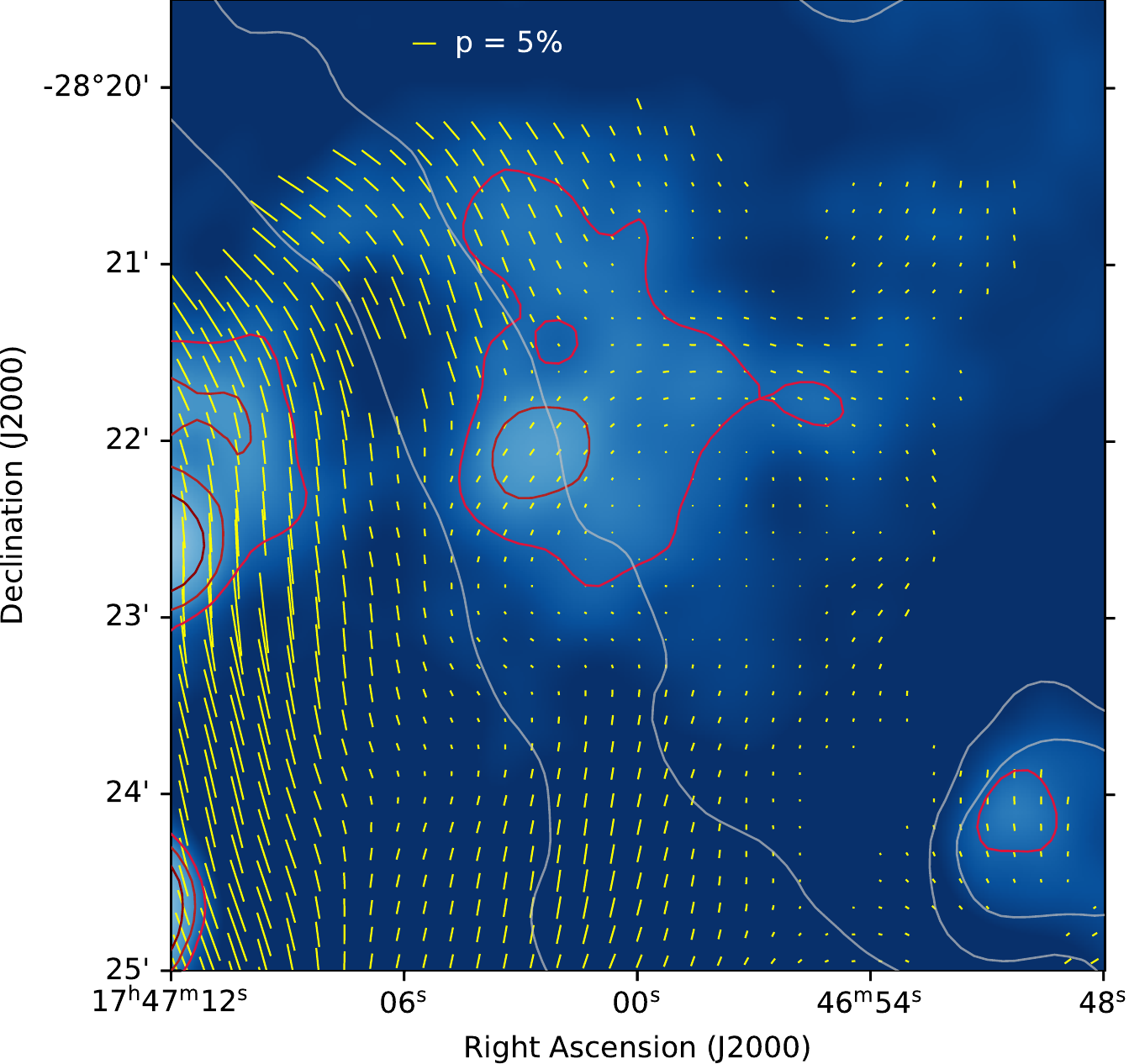}
\caption{
SgrB2 NW region (G0.65+0.03):  yellow pseudovectors show 214 \micron\ surface brightness above 5,000 \Mjsr overlaid on the 3 mm MGPS emission \citep[][smoothed to an angular resolution of 18\arcsec]{ginsburg20}. Red contours show the MGPS data at 10$\sigma$, 20$\sigma$, and 40$\sigma$ (for an rms value of 0.5 m\jybe). Grey contours show the MeerKAT 1 GHz data at 40\% and 50\% the peak emission \citep[][smoothed to the 18\arcsec\ resolution of the FIREPLACE survey]{heywood19}. 
}
\label{Sgb-nw}
\end{figure}

\subsection{Sgr B2 NW}

The Sgr B2 NW {(G0.65+0.03)} cloud is located northwest of the Sgr B2 Core \citep[see Figure 1 in][]{battersby20}. Figure \ref{Sgb-nw} shows the Sgr B2 NW cloud at 3 mm \citep{ginsburg20} overlaid with the 214 \micron\ polarization pseudovectors. This region has relatively low fractional polarization of $<$1\% compared with the other clouds included in this study. 
The cloud is located near the edge of the 1 GHz radio halo encompassing Sgr B2 and Sgr B1 (i.e., Sgr B complex, see Figures \ref{fig:LICs}, \ref{pol-fig}, and \ref{Sgb-nw}). However, we do not observe any spatial variation in the field direction or fractional polarization that could correlate with the radio halo in this vicinity, but instead observe stronger correlation between the pseudovectors and the 3 mm continuum morphology (Figure \ref{Sgb-nw}). This correlation is likely due to the MGPS and HAWC+ observations, at 3 mm and 214 \micron\ respectively, tracing the same dust continuum structures. 

In general, the brighter 3 mm emission regions tend to correspond to the lower fractional polarization regions. Regions South and East of the NW cloud, which contain lower 3 mm continuum emission, show higher fractional polarization values. It is generally only at the bright 3 mm peak in the cloud that we observe a slight increase in the fractional polarization values, when compared to neighboring pseudovectors (offset by 30--60\arcsec). The cloud does not appear to contain any compact 1 mm structures in the CMZoom survey \citep{Hatchfield20}, indicating that it may not yet be undergoing fragmentation.

\section{Conclusions}
\label{conclusion}

We present the first data release (DR1) of the Far-Infrared Polarimetric Large Area CMZ Emission (FIREPLACE) legacy survey (i.e., FIREPLACE I). FIREPLACE is a dust polarization survey of the entire CMZ to trace the magnetic fields in the cool dust component. The survey was taken with the 214 \micron\ (Band E) filter on the HAWC+ instrument aboard the SOFIA telescope. At these wavelengths we obtain an angular resolution of 19.6\arcsec\ (0.7 pc at the CMZ), which is roughly size of compact molecular clouds in this region. This DR1 covers roughly 1/3 of the CMZ and covers several prominent CMZ clouds including the Sgr B complex. We summarize the following scientific results from this first data release of the FIREPLACE Survey:

\begin{itemize}

\item \textbf{Two populations of Magnetic Field Orientations:} Analysis of the field directions over the entire coverage of the pilot study reveals a bimodal distribution of magnetic field directions (Section \ref{sec:histogram}). This bimodal distribution shows enhancements in the distribution of field directions for directions parallel and perpendicular to the Galactic plane (Figure \ref{hist-fig}). These two populations of field directions could be evidence for both poloidal and toroidal field components. Follow up analysis using the full FIREPLACE survey is needed to determine whether these two populations are present across the entire CMZ and how they compare with the DR1 distribution.

\item \textbf{Fractional polarization of CMZ clouds at 214 \micron:} 
We plot the fractional polarization against the 214 \micron\ intensity and find the data can be best fit with a slope of {-0.49 $\pm$0.01} (Figure \ref{PvI}).
We discuss the 214 \micron\ fractional polarization and magnetic field geometry of the area around Sgr B2 (Section \ref{sec:sources}), covering the following CMZ clouds: Sgr B2, Sgr B1, Dust Ridge clouds E/F, Sgr B2 NW, and Sgr B2 Halo (see Figure \ref{pol-fig} for identification of these sources). The \ring\ ring (M0.8-0.2) will be discussed in great detail in a follow-up publication: FIREPLACE II (Butterfield et al. 2023, submitted). The clouds observed in this DR1 pilot study have fractional polarization values around 1--5\%. Furthermore, the magnetic field generally follows the cloud morphology. 

\item \textbf{Low Fractional Polarization in Dense clouds:} The Sgr B2 Core and Cloud E/F show low fractional polarization towards the interiors of the clouds, where the column density is the highest. The low fractional polarization toward this region of the clouds is likely caused by some combination of the following three effects: 1) less grain alignment efficiency in the cloud interiors because of the attenuation of the radiation field there (RAT theory), 2) a varying polarization spectrum due to a superposition of multiple field directions along the line of sight, and 3) depolarization due to magnetic field entanglement below our resolution. 

\item \textbf{High Fractional Polarization towards \hii\ Regions:} 
We observe relatively high fractional polarization towards the regions surrounding several \hii\ regions, including the Sgr B2 sources V and R, G0.6-0.0, Sgr B1, and G0.73-0.10 (Figure \ref{pol-fig}). The more compact \hii\ regions (\eg, G0.6-0.0) appear to have higher fractional polarization (\til10\%) than their extended counterparts (\eg, Sgr B1, \til5\%). This could be a result of an enhancement in grain alignment in the dust surrounding the \hii\ regions, possibly caused by RAT alignment due to the enhanced radiation field from the O stars on which the \hii\ regions are centered.

\end{itemize}

The complete survey will cover the entire Galactic Center Central Molecular Zone (CMZ), roughly 1.5\degree\ in extent, from Sgr B2 to Sgr C. The second data release of the now-completed FIREPLACE survey (DR2) is in progress, with data in hand (FIREPLACE III, Par{\'e} et al., in prep). This complete survey will be invaluable for enhancing our big-picture perspective on the Galactic center magnetosphere.

\section*{Acknowledgements}

This work is based on observations made with the NASA/DLR Stratospheric Observatory for Infrared Astronomy (SOFIA). SOFIA is jointly operated by the Universities Space Research Association, Inc. (USRA), under NASA contract NNA17BF53C, and the Deutsches SOFIA Institut (DSI) under DLR contract 50 OK 2002 to the University of Stuttgart. Financial support for this work was provided by NASA through award \#09-0054 issued by USRA.

We would like to thank Dr. Simon Coude, Dr. Sachin Shenoy, Dr. Peter Ashton, Dr. Sarah Eftekharzadeh, Dr. Ryan Arneson, and the rest of the SOFIA team for their help with the observations and data reduction, including providing the 2.7.0 HAWC+ DRP software used in the reduction. 
We would like to thank Joe Michail (Northwestern) for his LIC python code used to create Figures  \ref{fig:LICs} and \ref{85rounds-lic} (based on IDL code by Diego Falceta-Gonçalves). The authors would also like to thank the anonymous referee for their helpful insight on this manuscript. 


\facility{SOFIA, MeerKAT, Herschel, GBT}

\software{\texttt{CRUSH}, \cite{Kovacs2008a}}

\appendix
\counterwithin{figure}{section}
\counterwithin{table}{section}

\renewcommand{\thesection}{A.\arabic{section}}
\renewcommand{\thefigure}{A.\arabic{figure}}
\addtocounter{figure}{-14}

\section{additional data reduction methodology}
\label{app-data-reduction}

\begin{figure*}[t!]
    \centering
    \includegraphics[height=0.95\textwidth]{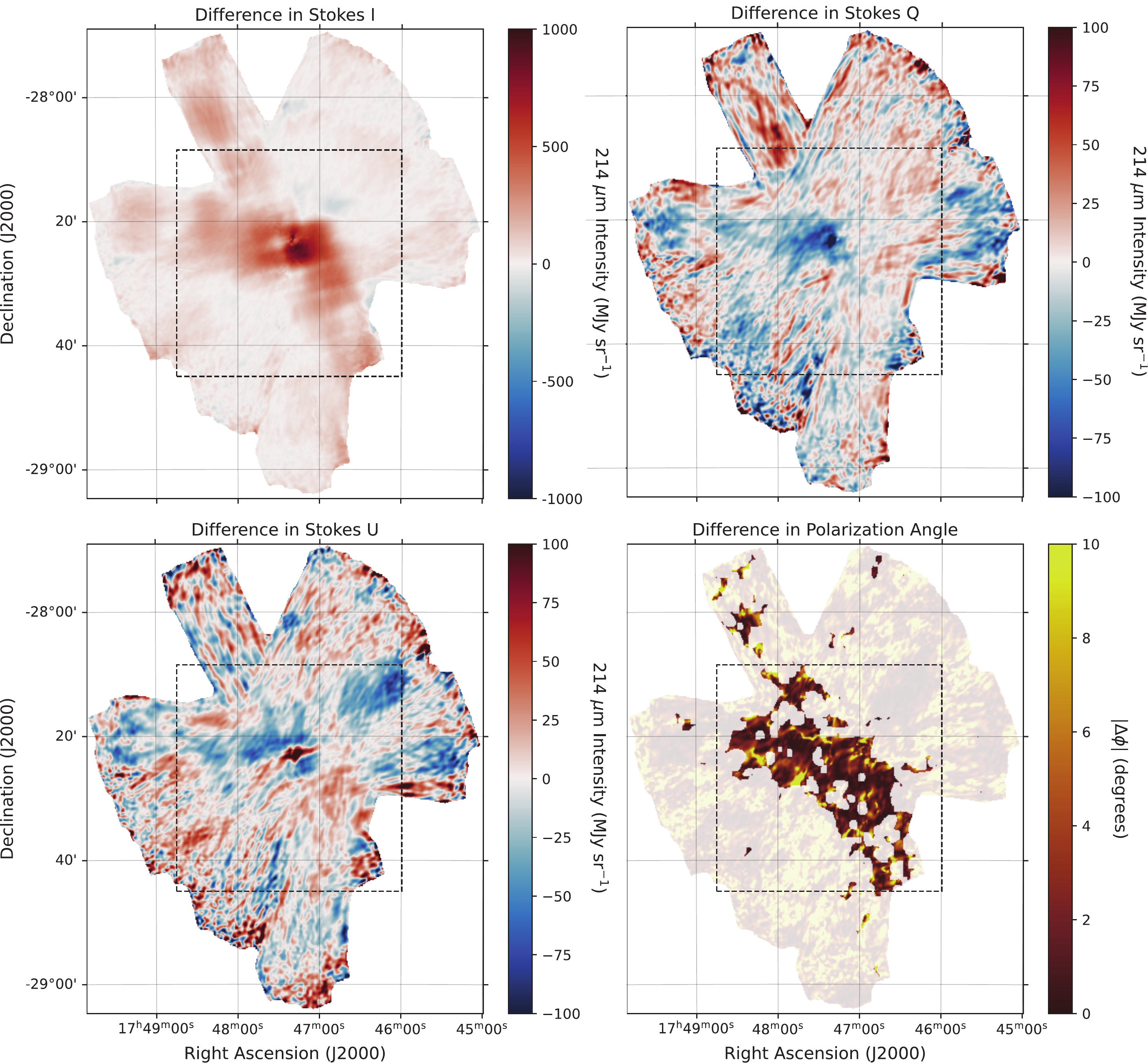}
    \caption{Maps of the differences in Stokes $I$ (top left), $Q$ (top right), $U$ (bottom left), and polarization angle (bottom right) for 75 and 85 iterations in the reduction pipeline are shown here. Such differences correspond to differences in polarization angle, $|\Delta \phi| \lesssim10^\circ$ across the region that passes the standard SOFIA cuts {(data points which do not satisfy the SOFIA cuts are shown as transparent in this panel)}. Because 10$^\circ$ is the statistical uncertainty associated with a 3$\sigma$ polarization measurements, we conclude that residuals are subdominant to statistical uncertainties. The black box shows the area containing the highest cross-linking (Figure \ref{hitmap}), and which corresponds to the field-of-view for Figures \ref{85rounds-crop}, \ref{fig:LICs}, \ref{hist-fig}, and \ref{pol-fig}.} 
    \label{fig:conv}
\end{figure*}

In this Appendix we discuss additional data reduction methodology that was not included in Section \ref{obs}. In Section \ref{sec:conv} we investigate the convergence of the data reduction by varying the ``\texttt{-rounds}'' parameter in \texttt{CRUSH}. In Section \ref{sec:sb2} we compare the magnetic field directions in Sgr B2 with previous CSO observations \citep{Dowell1998,Dotson10} to test for consistency between the observations. 
{Section \ref{app1.3} compares the scan-mode data, presented in Section \ref{survey}, with chop-nod data for Sgr B2 and Cloud E/F.}
In Section \ref{sec:Hcorr} we compare the Stokes $I$ intensity for each pixel with the Herschel 250 \micron\ observations \citep{molinari11} to also check for consistency between the observations. Lastly, we include a brief discussion on the line integral contours (LICs; Section \ref{sec:lic}) which are shown in several figures throughout this paper and are quite common in many polarization studies \citep[\eg,][]{Mangilli19}.

\subsection{Convergence of Reduction}
\label{sec:conv}

We tested the fidelity of the reduction pipeline using the parameters above by comparing our final reduction (\texttt{-rounds=85}) with a reduction using a lower number of iterations (\texttt{-rounds=75}).  These results are shown in Figure~\ref{fig:conv}.  Differences, between the two rounds of iterations, in the Stokes $Q$ and $U$ are within $\pm 100$ MJy sr$^{-1}$, with the exception of the brightest emission around the core of Sgr B2 (Figure \ref{fig:conv}). 
This leads to a polarization angle differences of $\lesssim 10^\circ$ across the part of the map that invokes the standard SOFIA cuts described in Section~\ref{obs}. Because 10$^\circ$ is the uncertainty in polarization angle that corresponds to a polarimetric signal-to-noise ratio of 3, residual variation due to variation in the number of iterations of the correlated noise fitting algorithm in our data reduction is subdominant to the statistical noise at our cut limit. This is shown in the lower right panel of Figure~\ref{fig:conv}. From this, we can conclude that the pipeline has converged to within satisfactory precision.

\subsection{Sgr B2 350~\micron\ Comparison}
\label{sec:sb2}

\begin{figure*}
    \centering
    \includegraphics[width=0.7\textwidth]{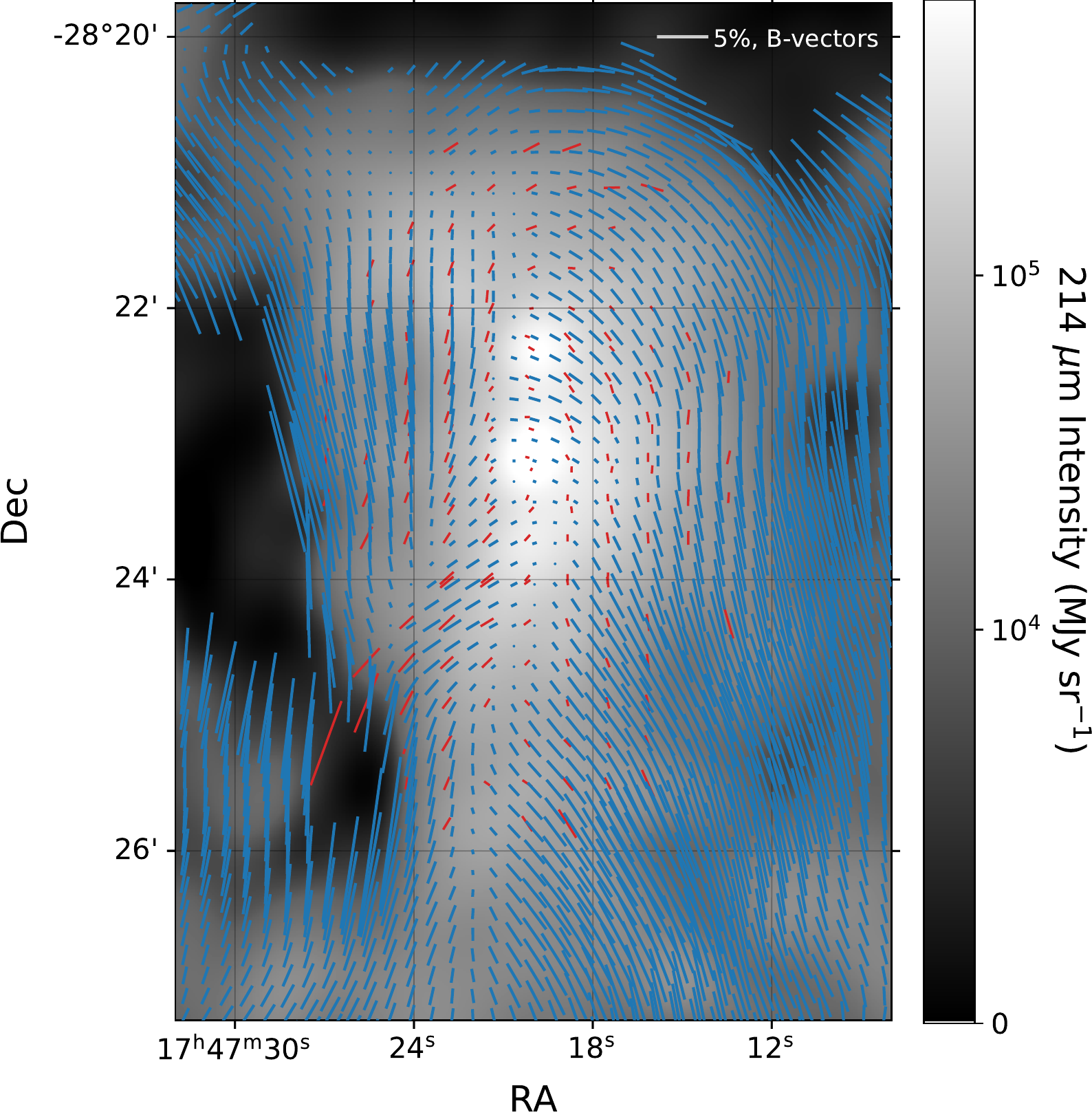}
    \caption{Inferred magnetic field pseudovectors toward Sgr B2 (scaled by polarization fraction) from HAWC+ at 214 \micron\ (blue) compared with the 350 \micron\ observations from Hertz/CSO \citep[red;][]{Dowell1998,Dotson10}.}  
    \label{fig:Sgrb2}
\end{figure*}

As a further check of our maps, we compare the measurements of the magnetic field toward Sgr B2 with previous (ground-based) polarimetric measurements at 350 \micron\ \citep{Dotson10}. Figure~\ref{fig:Sgrb2} shows reasonable agreement between the HAWC+ 214 \micron\ measurements reported here and the 350 \micron\ observations. The relatively small differences in the field orientation are quite possibly due to the well-known wavelength-dependence of polarization in this source \citep{Dowell1997,Novak1997}. This is due to the fact that the magnetic field {structure} in the hot cores differs from that in the cooler envelope. The superposition of these two polarization regimes along the line of sight produces variations in both the magnitude and direction of the polarization as a function of wavelength.

\subsection{Comparisons between Chop-nod and Scan-Mode Data}
\label{app1.3}

\begin{figure*}
    \centering
    \includegraphics[width=\textwidth]{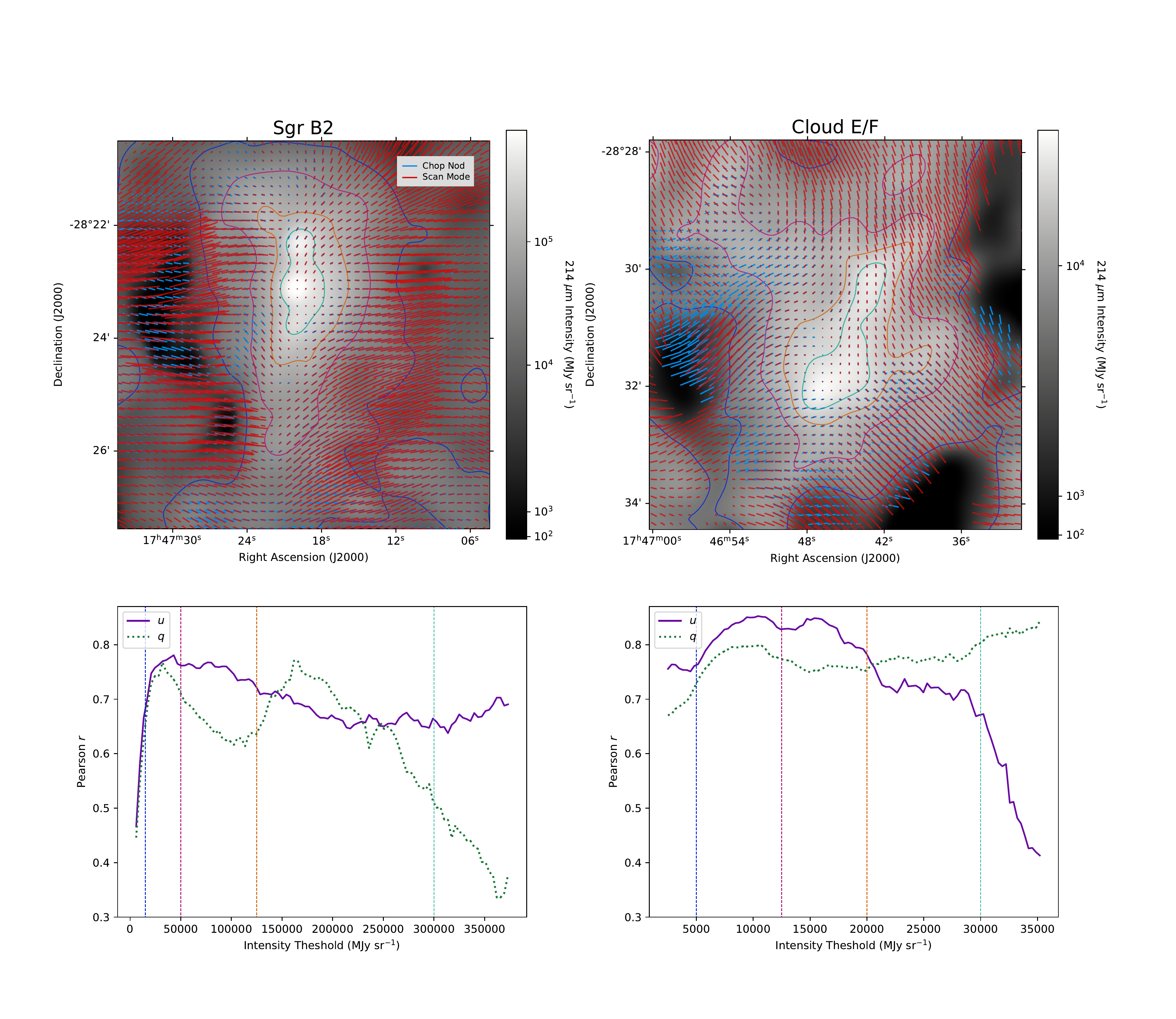}
    \caption{(Top) Map-domain comparisons between the chop nod (blue) and scan mode (red) {electric field} (polarization direction) pseudovectors in Sgr B2 and Cloud E/F are shown. (Bottom) For each cloud, we show the Pearson $r$ coefficient between the normalized Stokes parameters, $u$ (solid purple) and $q$ (dotted green) of the chop nod and scan mode data as a function of intensity threshold. The correlation is calculated between points where the total intensity exceeds the intensity threshold.}  
    \label{fig:cn}
\end{figure*}

{To further validate the scan-mode data, we obtained chop-nod data for the Sgr B2 and Cloud E/F regions (see Figure \ref{pol-fig}, for the locations of these clouds), using the SOFIA/HAWC+ instrument. The chop nod data were obtained on June 23, 2022 during flight 890. The Sgr B2 data consist of 7 files and the Cloud E/F data consist of 4 files.}

{Figure ~\ref{fig:cn} (top) shows the inferred {electric field} pseudovectors of the chop-nod and scan-mode data in the Sgr B2 and Cloud E/F regions. A polarimetric signal-to-noise cut, $p/\sigma_p>3$, and a total intensity signal-to-noise cut, $I/\sigma_I>200$, were applied. Less agreement is seen around low intensity regions where the reference beam contamination is more significant and there are lower signal-to-noise ratios.} 

{Figure ~\ref{fig:cn} (bottom) shows the Pearson $r$ correlation between the normalized Stokes of the chop nod and scan mode data as a function of intensity threshold. The Pearson $r$ value is calculated for all points above the threshold intensity, as calculated from the scan-mode Stokes $I_{214\mu m}$ map. The maximum intensity value was chosen to ensure that a minimum of 100 pixels (approximately 6 independent beams) is included in each correlation.}

{For low values of threshold, the correlations are observed to drop. This is likely due to map filtering differences; e.g., reference beam contamination for the chop-nod mode in low intensity regions is expected to affect the measurement. For high thresholds, loss of correlation is observed in the normalized Stokes parameter that has the lower value.  This is likely due to a lower-signal-to noise for that parameter. In general, the agreement between the two maps is good, supporting the validity of the data reduction pipeline.}

\subsection{Herschel Stokes $I$ Comparison}
\label{sec:Hcorr}

\begin{figure*}
    \centering
    \includegraphics[width=0.89\textwidth]{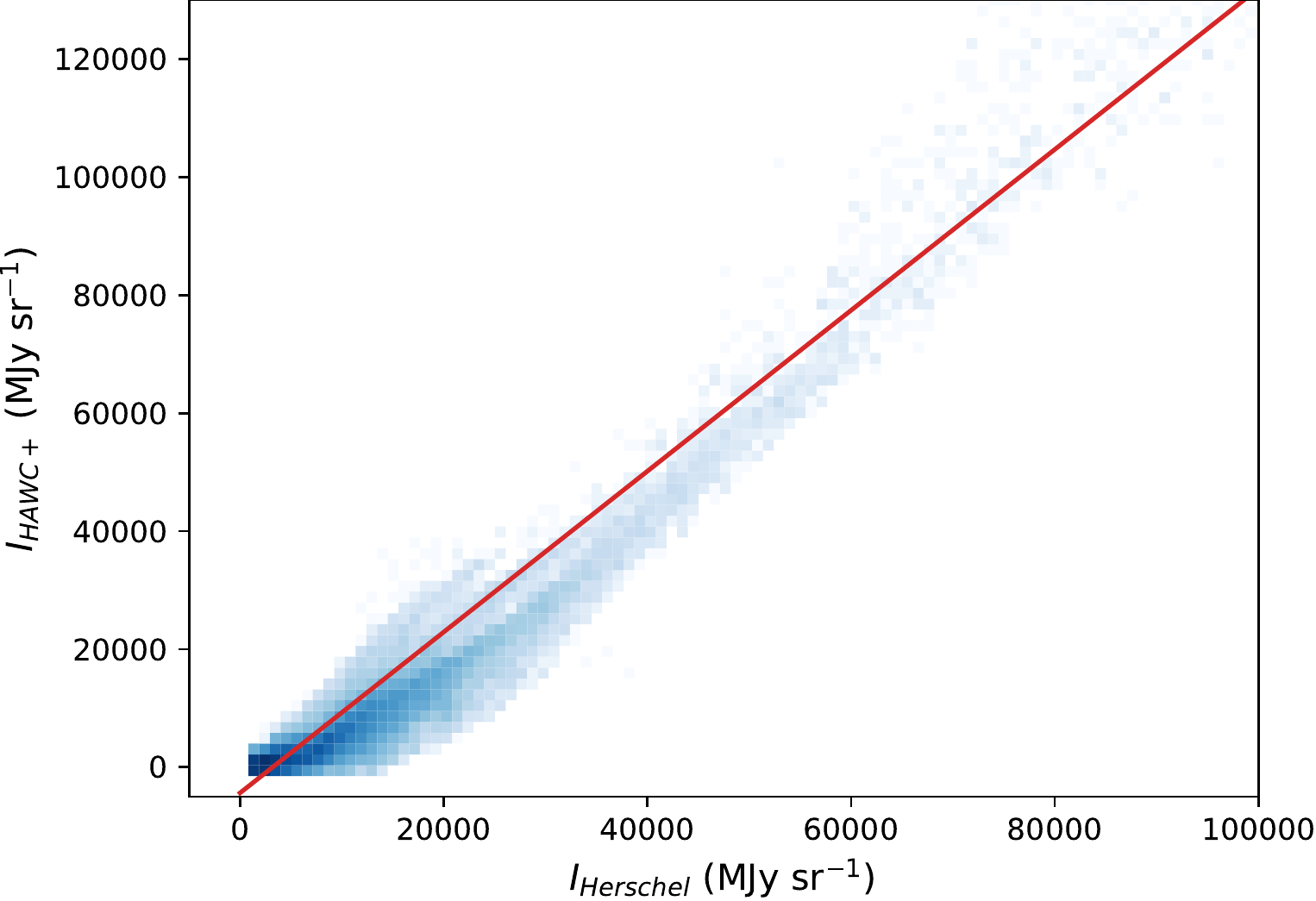}
    \caption{The correlation in Stokes $I$ between the SOFIA/HAWC+ 214~\micron\ and Herschel/SPIRE 250~\micron\ maps is shown as a two-dimensional histogram. The Pearson $r$ value is found to be 0.95 and the best fit slope \textbf{(red line)} between the data sets is 1.36, consistent with characteristic dust temperatures and spectral index values in the GC. 
    }
    \label{fig:hcorr}
\end{figure*}

We also confirm that our Stokes $I$ map is in agreement with the 250 \micron\ Herschel/SPIRE map of the region \citep{molinari11}. A two-dimensional histogram comparing the FIREPLACE DR1 region with the same region in the Herschel data is shown in Figure~\ref{fig:hcorr}.  The Pearson $r$ correlation coefficient is 0.95, indicating good agreement.  The slope of the best-fit line is 1.36.  This is a reasonable color-correction factor between 214 and 250 \micron. For example, this is consistent with a a greybody having $\beta=2$ and $T=22$ K, reasonable representative parameters for cool dust in the CMZ region. 

\subsection{Line Integral Contour Map}
\label{sec:lic}

\begin{figure*}
\centering
\includegraphics[width=1.0\textwidth]{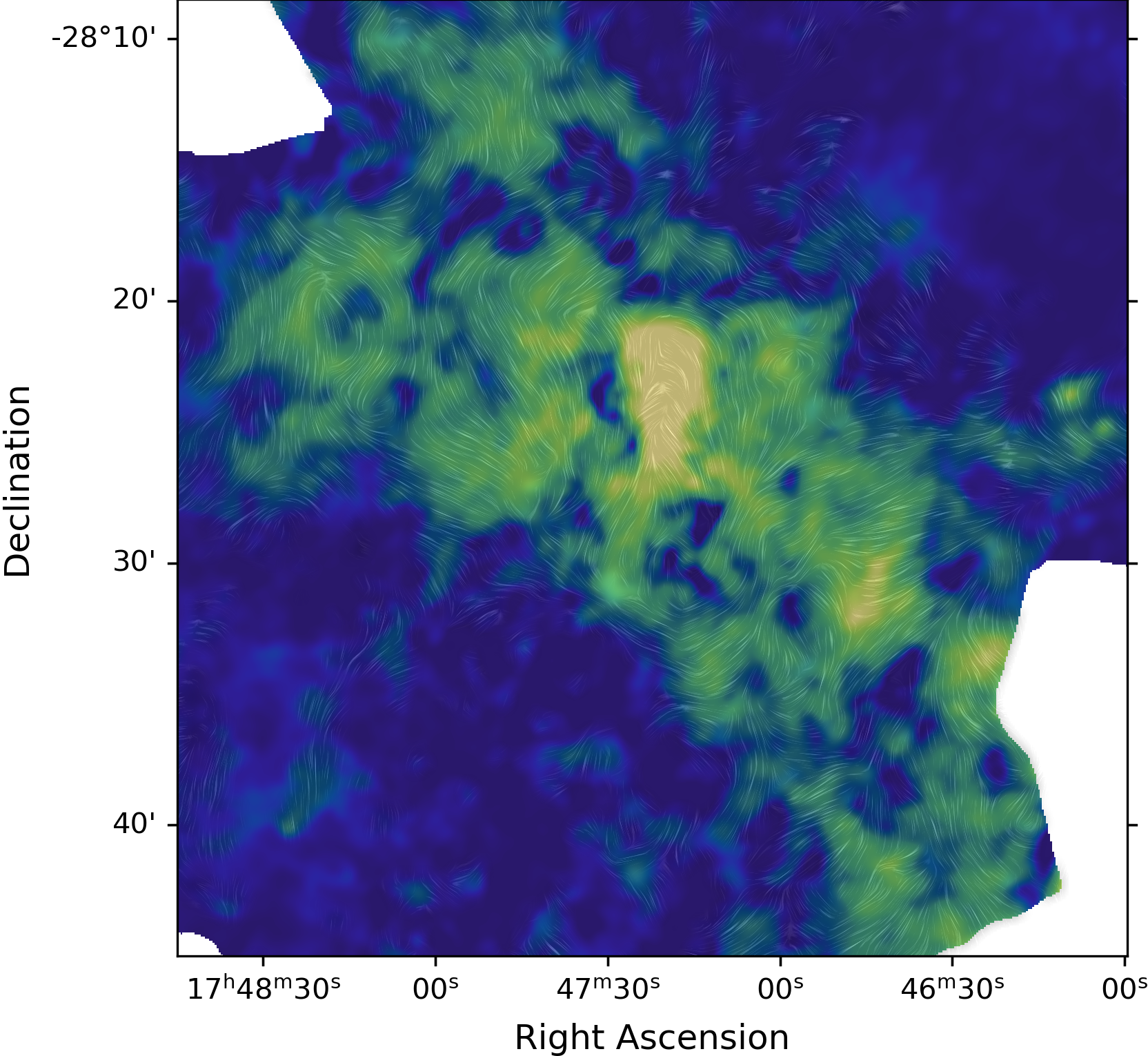}
\caption{Line integral contour \citep[LIC;][]{Cabral1993} representations of the magnetic field in the cool dust superimposed on the 214 \micron\ SOFIA total intensity map. A version of this Figure showing the corresponding magnetic field pseudovectors is shown in Figure \ref{85rounds-crop}. The LICs shown in this figure are also overlaid on the Herschel 250 \micron\ and MeerKAT 1 GHz datasets in Figure \ref{fig:LICs}.  
}
\label{85rounds-lic}
\end{figure*}

A line integral contour (LIC) is a method of illustrating the polarization vectors as stream lines \citep[][]{Cabral1993}. This method filters the input array of polarization pseudovectors along local stream lines to produce an output image that represents the `flow' direction of the lines. This visualization method can be a beneficial way of viewing the data as it can help illustrate large scale field directions and is a common way of presenting polarization data \citep[\eg,][see their Figure 1]{Mangilli19}. Figure \ref{85rounds-lic} shows the LIC map of the data shown in Figure \ref{85rounds-crop}. These LICs are also overlaid on the Herschel 250 \micron\ and MeerKAT 1 GHz datasets in Figure \ref{fig:LICs}. 

\bibliographystyle{aasjournal}
\bibliography{main.bib}

\end{document}